%% file: main.tex
\documentclass[a4paper,11pt]{article}
\pdfoutput=1

\usepackage{jinstpub}
\usepackage{siunitx}

\usepackage{xspace}
\usepackage{booktabs}

\usepackage{subcaption}

\title{\boldmath EUDAQ -- A Data Acquisition Software Framework for Common Beam Telescopes}

\input{authors}

\input{abstract}

\arxivnumber{1909.13725} 

\input{definitions.tex}

\begin{document}
\maketitle
\flushbottom


\input{content/intro}

\input{content/framework}

\input{content/usage}

\input{content/application}


\input{content/conclusion}

\input{acknowledgment}

\bibliographystyle{elsarticle-num}
\bibliography{references}

\end{document}

%% file: authors.tex
\author[c]{P.~Ahlburg,}
\author[b, \textasteriskcentered]{S.~Arfaoui, \note[\textasteriskcentered]{Deceased.}}
\author[a]{J.-H.~Arling,}
\author[d]{H.~Augustin,}
\author[b]{D.~Barney,}
\author[e]{M.~Benoit,}
\author[f]{T.~Bisanz,}
\author[e,s]{E.~Corrin,}
\author[g]{D.~Cussans,}
\author[b]{D. Dannheim,}
\author[a,2]{J.~Dreyling-Eschweiler, \note{Corresponding author: jan.dreyling-eschweiler@desy.de}}
\author[a]{T.~Eichhorn,}
\author[b,t]{A.~Fiergolski,}
\author[a]{I.-M.~Gregor,}
\author[f]{J.~Grosse-Knetter,}
\author[e,u]{D.~Haas,}
\author[d]{L.~Huth,}
\author[a,h]{A.~Irles,}
\author[a]{H.~Jansen,}
\author[c]{J.~Janssen,}
\author[b]{M.~Keil,}
\author[a,i]{J.~S.~Keller,}
\author[d,e]{M.~Kiehn,}
\author[b]{H.~J.~Kim,}
\author[j]{J.~Kroll,}
\author[a]{K.~Kr\"uger,}
\author[b]{S.~Kulis,}
\author[a,j]{J.~Kvasnicka,}
\author[f]{J.~Lange,}
\author[a]{Y.~Liu,}
\author[c]{F.~L\"utticke,}
\author[c,q]{C.~Marinas,}
\author[b]{P.~Martinengo,}
\author[p]{A. Nurnberg,}
\author[c]{B.~Paschen,}
\author[a,k]{H.~Perrey,}
\author[a,l]{R.~Peschke,}
\author[a]{D.~Pitzl,}
\author[c]{D.-L.~Pohl,}
\author[f]{A.~Quadt,}
\author[b,r]{T.~Quast,}
\author[b]{F.~Reidt,}
\author[a]{E.~Rossi,}
\author[a,v]{I.~Rubinsky,}
\author[b]{A.~Rummler,}
\author[f]{H.~Schreeck,}
\author[a]{P.~Sch\"utze,}
\author[f]{B.~Schwenker,}
\author[b]{S.~Spannagel,}
\author[a]{M.~Stanitzki,}
\author[f]{U.~Stolzenberg,}
\author[m]{T.~Suehara,}
\author[b]{M.~Suljic,}
\author[n,w]{G.~Troska,}
\author[o]{M.~Varga-Kofarago,}
\author[f,n]{J.~Weingarten,}
\author[f]{and P.~Wieduwilt}

\affiliation[a]{Deutsches Elektronen-Synchrotron (DESY),\\Notkestr. 85, 22607 Hamburg, Germany}
\affiliation[b]{European Organization for Nuclear Research (CERN),\\1211 Geneva 23, Switzerland}
\affiliation[c]{Physikalisches Institut, Universit\"at Bonn,\\ Nu\ss allee 12, 53115 Bonn, Germany}
\affiliation[d]{Physikalisches Institut, Universit\"at Heidelberg,\\Im Neuenheimer Feld 226, 69120 Heidelberg, Germany}
\affiliation[e]{D\'epartement de physique nucl\'eaire et corpusculaire, Universit\'e de Gen\`eve\\24, quai Ernest-Ansermet, 1211 Gen\`eve, Switzerland}
\affiliation[f]{II. Physikalisches Institut, Georg-August-Universit\"at G\"ottingen,\\Friedrich-Hund-Platz 1, 37077 G\"ottingen, Germany}
\affiliation[g]{University of Bristol, H.H. Wills Physics Laboratory,\\Tyndall Avenue, Bristol BS8 1TL, United Kingdom}
\affiliation[h]{Laboratoire de l'Acc\'elerateur Lin\'eaire (LAL), CNRS/IN2P3 et Universit\'e de Paris-Sud XI,\\Centre Scientifique d'Orsay, B\^atiment 200, BP 34, F-91898 Orsay, CEDEX, France}
\affiliation[i]{Physics Department, Carleton University, 1125 Colonel By Drive, Ottawa, Ontario, K1S 5B6, Canada}
\affiliation[j]{Institute of Physics of the Czech Academy of Sciences,\\Na Slovance 2, 18221 Prague 8, Czech Republic}
\affiliation[k]{Division of Nuclear Physics, Lund University,\\SE-221 00, Lund, Sweden}
\affiliation[l]{Department of Physics and Astronomy, University of Hawai'i at M\={a}noa, \\Watanabe 416, 2505 Correa Road, Honolulu, HI 96822, USA}
\affiliation[m]{Department of Physics, Kyushu University,\\744 Motooka, Nishi-ku, Fukuoka 819-0395, Japan}
\affiliation[n]{Experimentelle Physik IV, Technische Universit\"at Dortmund,\\Otto-Hahn-Str. 4A, 44227 Dortmund, Germany}
\affiliation[o]{Hungarian Academy of Sciences Wigner Research Centre for Physics,\\Konkoly-Thege Mikl\'os \'ut 29-33, 1121 Budapest, Hungary}
\affiliation[p]{Karlsruhe Institute of Technology (KIT), IPE, 76021 Karlsruhe, Germany}
\affiliation[q]{Instituto de Fisica Corpuscular (IFIC), University of Valencia,\\Catedratico Jose Beltran 2, E-46980 Paterna, Spain}
\affiliation[r]{Physikalisches Institut 3A, RWTH Aachen,\\Otto-Blumenthal-Str., 52074 Aachen, Germany}

\affiliation[s]{now at: Microsoft UK,\\2 Kingdom Street, Paddington, London W2 6BD}
\affiliation[t]{now at: Fastree3d,\\EPFL Innovation Park,  B\^atiment L, 1B, Chemin de la Dent d'Oche, 1024 Ecublens, Switzerland}
\affiliation[u]{now at: Airbus Defence and Space GmbH,\\Claude-Dornier-Str., 88090 Immenstaad, Germany}
\affiliation[v]{now at: d-fine GmbH,\\Bavariafilmplatz 8, 82031 Gr\"unwald,  Germany}
\affiliation[w]{now at: Infineon Technologies AG,\\Max-Planck-Str. 5, 59581 Warstein, Germany}

%% file: abstract.tex
\abstract{
EUDAQ is a generic data acquisition software developed for use in conjunction with common beam telescopes at charged particle beam lines.
Providing high-precision reference tracks for performance studies of new sensors, beam telescopes are essential for the research and development towards future detectors for high-energy physics.
As beam time is a highly limited resource, EUDAQ has been designed with reliability and ease-of-use in mind.
It enables flexible integration of different independent devices under test via their specific data acquisition systems into a top-level framework.
EUDAQ controls all components globally, handles the data flow centrally and synchronises and records the data streams.
Over the past decade, EUDAQ has been deployed as part of a wide range of successful test beam campaigns and detector development applications.
}

%% file: definitions.tex
\newcommand{\desyii}{{\mbox{DESY II}}\xspace}
\newcommand{\diitbf}{{\desyii Test Beam Facility}\xspace}
\newcommand{\geant}{\textsc{Geant4}\xspace}
\newcommand{\EUDAQ}{\textsc{EUDAQ}\xspace}
\newcommand{\EUDAQII}{\textsc{EUDAQ2}\xspace}
\newcommand{\EUTELESCOPE}{\textsc{EUTelescope}\xspace}
\newcommand{\python}{\textsc{PYTHON}\xspace}
\newcommand{\posix}{\textsc{POSIX}\xspace}
\newcommand{\ROOT}{\textsc{ROOT}\xspace}
\newcommand{\LCIO}{\textsc{LCIO}\xspace}
\newcommand{\CMAKE}{\textsc{CMake}\xspace}

\newcommand{\MIMOSA}{{Mimosa26}\xspace}

\newcommand{\EUDET}{\textsc{EUDET}\xspace}
\newcommand{\AIDA}{\textsc{AIDA}\xspace}
\newcommand{\AIDAII}{\textsc{AIDA-2020}\xspace}

\newcommand{\RunControl}{\textit{Run\,Control}\xspace}
\newcommand{\Producer}{\textit{Producer}\xspace}
\newcommand{\Producers}{\textit{Producers}\xspace}
\newcommand{\DataCollector}{\textit{Data\,Collector}\xspace}
\newcommand{\Storage}{\textit{Storage}\xspace}
\newcommand{\LogCollector}{\textit{Log\,Collector}\xspace}
\newcommand{\LogSender}{\textit{Log\,Sender}\xspace}
\newcommand{\LogMessage}{\textit{Log\,Message}\xspace}
\newcommand{\OnlineMonitor}{\textit{Online\,Monitor}\xspace}
\newcommand{\NIProducer}{\textit{NIProducer}\xspace}
\newcommand{\TLUProducer}{\textit{TLUProducer}\xspace}
\newcommand{\SlowProducer}{\textit{Slow\,Producer}\xspace}
\newcommand{\SlowProducers}{\textit{Slow\,Producers}\xspace}
\newcommand{\MagicLogBook}{\textit{Magic\,Log\,Book}\xspace}

\newcommand{\Event}{\textit{Event}\xspace}
\newcommand{\RAWDATAEVENT}{\textit{RawDataEvent}\xspace}
\newcommand{\DATACONVERTERPLUGIN}{\textit{Data\,ConverterPlugin}\xspace}
\newcommand{\DATACONVERTERPLUGINS}{\textit{DataConverterPlugins}\xspace}
\newcommand{\SERIALIZABLE}{\textit{Serializable}\xspace}
\newcommand{\COMMANDRECEIVER}{\textit{CommandReceiver}\xspace}
\newcommand{\DATASENDER}{\textit{DataSender}\xspace}
\newcommand{\Serialize}{\textit{Serialize}\xspace}
\newcommand{\Serializer}{\textit{Serializer}\xspace}
\newcommand{\Deserializer}{\textit{Deserializer}\xspace}
\newcommand{\SendEvent}{\textit{SendEvent}\xspace}

\newcommand{\ONINITIALISE}{\textit{OnInitialise}\xspace}
\newcommand{\ONCONFIGURE}{\textit{OnConfigure}\xspace}
\newcommand{\ONSTARTRUN}{\textit{OnStartRun}\xspace}
\newcommand{\ONSTOPRUN}{\textit{OnStopRun}\xspace}

\DeclareSIUnit\neq{\text{$n_\textup{eq}$}\per\cm\squared}

%% file: content/intro.tex
\section{Introduction}
\label{sec:intro}
Test beam campaigns with beam telescopes constitute a crucial component in the research and development of novel particle detectors.
In-depth performance studies usually require simultaneous operation of different detector systems and synchronised recording of data.
This often poses a significant challenge due to the necessary hardware and software integration into a common data acquisition (DAQ) system.

\EUDAQ is a generic DAQ software framework that was developed to simplify this integration process.
Written in C++, supporting multiple platforms such as Linux, MacOS~X and Windows and distributed under the LGPLv3~\cite{LGPLv3}, \EUDAQ provides a high level of interoperability.
As such, it is one of the few examples of common tools for test beams today and is well-received by a growing user community~\cite{bttb6forum}. 

The development of \EUDAQ started as part of the EU-funded Joint Research Activity of the \EUDET project ``Test Beam Infrastructure'' \cite{eudet-www, eudet2012} in 2005.
The goal of the project was the development of a high-precision pixel beam telescope for investigations of the tracking performance of sensor devices.
To make this beam telescope a versatile tool for a broad user base and a wide variety of devices, an easy integration strategy for devices under test (DUT) and its DAQ was a priority from the beginning of the project~\cite{haas2006}. 
The interface that was considered to be the most flexible for the user consisted of two separate layers:
on the hardware level, the different DAQ systems were to be synchronised using a simple trigger-busy communication protocol;
on the software level, the full integration of the DUT into \EUDAQ  was foreseen as the preferred approach, yet was kept optional.

This approach determined the core architecture of \EUDAQ~\cite{corrin2010} which remains today:
centralised but distributed core components that communicate with so-called \Producers via a custom TCP/IP-based protocol.
In this scheme, the latter are responsible for implementing an interface to the individual hardware components, controlling the devices' states and feeding the data into the central data collection unit.
Both the beam telescope detector planes and the DUT use the same interface, thus making the framework flexible and independent of any specific hardware.
The framework architecture is described in more detail in Section~\ref{sec:framework}.

Historically, the most prominent application of \EUDAQ is the DAQ of the \EUDET-type pixel beam telescopes~\cite{jansen2016}.
They are based on \MIMOSA sensors \cite{huguo2010} as telescope planes and a custom-designed trigger logic unit (TLU), the \EUDET TLU \cite{tlu}. 
Today, the \EUDET-type beam telescopes are accessible as common infrastructure at test beam facilities all over the world.
This broad availability of beam telescopes combined with the ease-of-use, extensive documentation and user-focus of \EUDAQ outlined in Section~\ref{sec:usage} has led to a large number of successful \EUDAQ-based test beam campaigns in the last decade:
in Section~\ref{sec:application}, eleven applications from a wide range of communities are described in more detail.

Since the early days of mostly user-driven development, \EUDAQ has been moved to a collaborative development model with several active contributors.
New features as well as many behind-the-scenes changes such as continuous integration methods paved the road towards the second major version of \EUDAQ as  briefly outlined in Section~\ref{sec:conclusion}.

%% file: content/framework.tex
\section{Framework Architecture}
\label{sec:framework}

\begin{figure}[htbp]
\begin{center}
\includegraphics[width=0.8\textwidth]{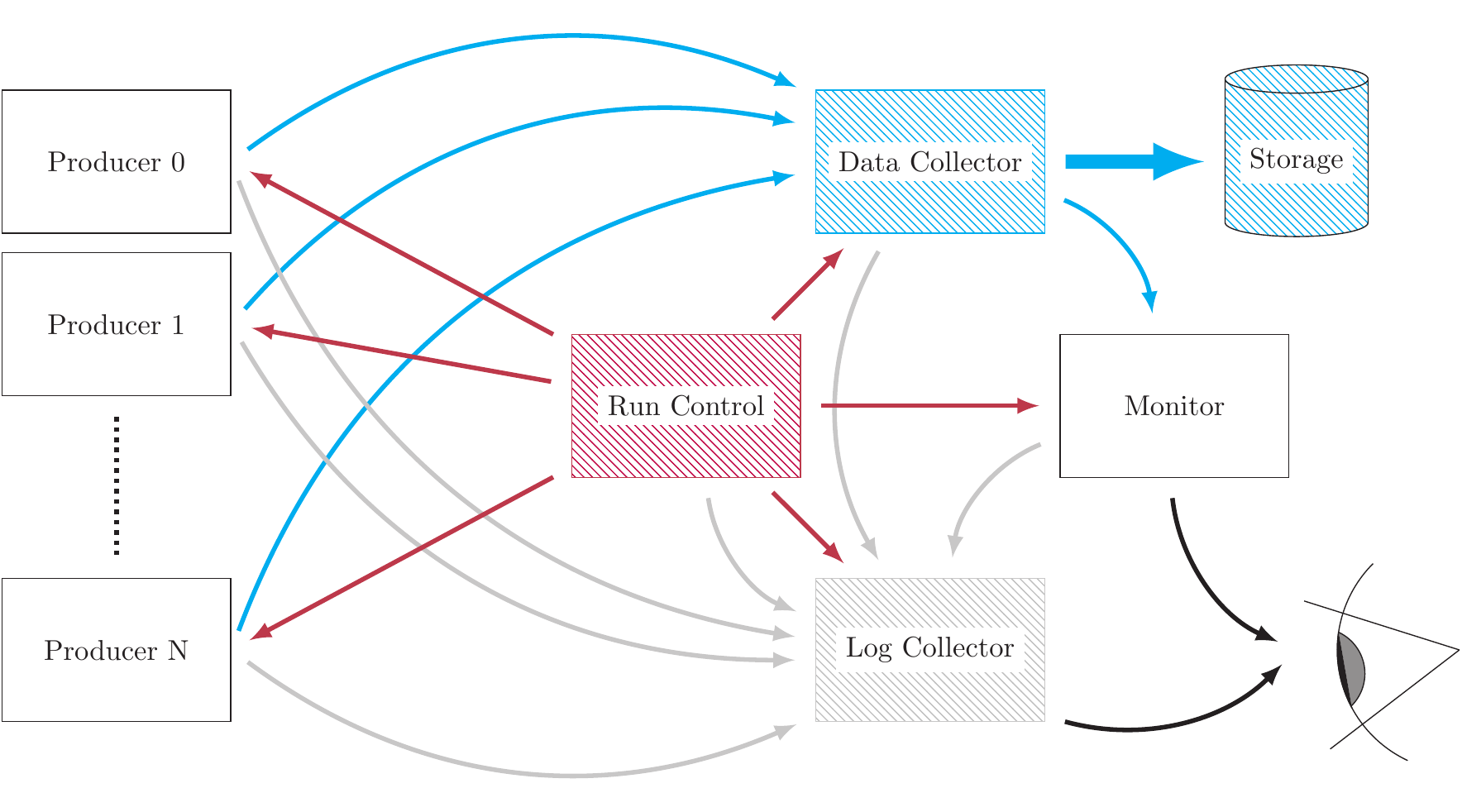}
	\caption{Sketch of the \EUDAQ architecture depicting the core components and the data flow. The individual communication channels are indicated by arrows: the control channel (red), the data channel (blue) and the log channel (grey). Figure from \cite{cmspixel-spannagel}.
    }
\label{fig:framework:sketch}
\end{center}
\end{figure}

The \EUDAQ framework is split into a number of different processes,
usually running on several different computers connected via LAN, and communicating using a custom TCP/IP protocol, indicated by arrows between the components in Figure~\ref{fig:framework:sketch}.
The framework is written in C++, with minimal external dependencies, in order to ensure maximal portability.
It uses \posix threads for concurrency and \posix sockets~\cite{8277153} for inter-process communication over TCP/IP.

There are three separate channels for sending different types of information to the individual \EUDAQ processes: the command, the data and the log channel.
In the following, a brief overview of the framework architecture and its concepts is given in Section~\ref{sec:framework:overview}, and the individual communication channels are introduced in more detail in Sections~\ref{sec:framework:command},~\ref{sec:framework:data}~and~\ref{sec:framework:log}.
The architecture overview is complemented by a description of the online monitoring tool in Section~\ref{sec:framework:onlinemon} as a crucial part of the user interface.

\subsection{Overview \& Concepts}
\label{sec:framework:overview}

A central \RunControl provides an interface to the user for controlling the DAQ system.
The \RunControl monitors the state of all other processes and sends them commands to start and stop a run, configure the components etc. using the command channel indicated by the red arrows in Figure~\ref{fig:framework:sketch}.
Here, \emph{run} indicates a continuous period of data taking with typical durations ranging from a few minutes to several hours.
Runs are divided into so-called \emph{events}, which represent the data from all participating devices corresponding to one trigger.
Triggers are synchronised between different devices at the hardware level.

Each process of the DAQ that produces data is controlled by a \Producer.
The \EUDAQ \Producer class provides a standardised interface to the rest of the framework for receiving commands and sending data and log messages.
A \Producer should send one event for every trigger.
Each event consists of a standard header containing the run number, the trigger number and some other status information, along with the unprocessed raw data from the sensor.
If a \Producer has no data for certain events for example a small sensor that does not register a hit for every trigger, then it should send empty events (see Section~\ref{sec:application:mu3e}).
In addition to these standard events, each \Producer must also send special beginning-of-run and end-of-run events at the start and end of each run respectively.
The beginning-of-run event includes the configuration data for the run, ensuring that this information is always available with the collected data.
For devices that generate data much less frequently or triggerless data, for example temperature sensors, motor stages (see Section \ref{sec:application:hgcal} or \ref{sec:application:mbi})
or other environment monitors, \EUDAQ provides an alternative base class called \SlowProducer.
However, it is encouraged to use the standard \Producers whenever feasible, since the strong synchronisation of events provided helps to ensure data integrity.

The \DataCollector receives data as \EUDAQ \Event objects from all data-generating processes over the data channel as shown by the blue arrows in Figure~\ref{fig:framework:sketch}.
It collates and organises them such that event data from all \Producers is ordered and synchronised, and then for each trigger it combines all the relevant events and writes them to disk (\Storage in Figure~\ref{fig:framework:sketch}).
Because all \Producers send one event for every trigger, along with the beginning and end of run events to mark the runs,
the synchronisation between \Producers to be kept simple, and any problems can be caught and rectified as early as possible.
During event building the \DataCollector ignores the absence of events from \SlowProducers,
which are distinguished from the usual \Producers by the type information that is communicated during the initial connection.
By default, the data files are just the serialised \EUDAQ \Event objects, consisting of a common header with run number, trigger number etc. and the raw binary data provided by the respective detector.
By directly recording the raw detector response without online processing, all initially available information is retained and the risk posed by possible mistakes in the intermediate data processing code is mitigated.
These architectural decisions have proven to provide a high degree of stability and confidence in the integrity of the collected data.
If required, \EUDAQ may also be configured to write in various other formats, such as the \LCIO format~\cite{Gaede:2003ip,Aplin:2012kj} or its own \emph{StandardEvent} format.

Log messages from all processes are collected centrally in the \LogCollector, using the dedicated log channel indicated by gray arrows in Figure~\ref{fig:framework:sketch}.
This provides a central location for the user to monitor all distributed processes for any unexpected warnings or errors.
The \OnlineMonitor reads the output data file and generates a range of plots to help the user verify that the system is operating properly.

Data is serialised using a simple \EUDAQ-specific binary format.
The primitive types like different-sized integers, floats etc. are serialised as a sequence of bytes using little-endian format even on big-endian CPU architectures, so that the data is compatible between different machine architectures.
Strings are serialised using a 32-bit length indicator followed by the contents of the string interpreted as bytes. A range of standard types, such as \texttt{std::vector} and \texttt{std::map} are also handled by \EUDAQ, using a similar scheme.
Classes defined in the \EUDAQ code base, or by the user, are made serializable by inheriting from the \SERIALIZABLE base class, and by implementing the virtual \Serialize function and a constructor that takes a single \Deserializer instance.
In most cases, the implementation is as simple as writing each member variable to the \Serializer in the \Serialize method, and reading them
in the same order in the constructor.
All member variables of a \SERIALIZABLE class must also be serializable -- either by inheriting from \SERIALIZABLE, or being explicitly handled
by the \Serializer class.

\subsection{Command Channel}
\label{sec:framework:command}

The command channel is controlled by the \RunControl component.
All other components are connected to the command channel in order to keep track of the current state of the DAQ system.
Components that communicate with the \RunControl must inherit from the \COMMANDRECEIVER class, which handles the reception of the command messages and calls the corresponding member functions.
When receiving a command from the finite state machine of the \RunControl, the component must execute the requested state transition and respond with its new state.

This allows the \RunControl to monitor the states of individual components, and to define the global machine state.
Messages on the command channel are kept simple, consisting of a single string command with an optional string parameter.
This is encoded into a single string message, separated by a null byte if the optional parameter is present.
The individual commands are listed in Table~\ref{tab:framework:CommandChannel}.

\begin{table}
\caption{\label{tab:framework:CommandChannel} The individual commands defined for distribution via the \EUDAQ command channel}
\begin{center}
 \begin{tabular}{ l l l }
    \toprule
   \bfseries Command   & \bfseries Parameter     & \bfseries Description \\
   \midrule
   INIT      & Configuration & Initialise DAQ system \\
   CONFIG    & Configuration & Configure DAQ for a run \\
   START     & Run number    & Start a new run (and send run number) \\
   STOP      & n/a           & Stop a run \\
   RESET     & n/a           & Reset DAQ system \\
   TERMINATE & n/a           & Terminate (quit) DAQ \\
   STATUS    & n/a           & Retrieve status of all systems \\
   LOG       & Log server    & Configure log server address \\
   \bottomrule
 \end{tabular}
 \end{center}
\end{table}

\subsection{Data Channel}
\label{sec:framework:data}

The data channel is used to collect all data generated during a run
into a single component, where it is written to disk.
This component is called \DataCollector and acts as receiver for data from any of the \Producers participating in the run.
Usually this will include the \Producer for the telescope DAQ, a
\Producer for the DUT DAQ, and a \Producer for a system providing trigger information.
All \Producers inherit from the \Producer base class, which provides a
\SendEvent method that the inheriting class should call for each event to be sent to permanent storage.

%

\subsection{Log Channel}
\label{sec:framework:log}

At runtime, every process can generate log
messages containing information about the state of the component, or
any unexpected occurrences.
Simple access to the messages during data taking is realised by a central \LogCollector,
which collects messages from all Log channels.

Every process inherits from the \LogSender class, enabling it to send log messages over the TCP/IP connection to the \LogCollector.
Each log message is an instance of the \LogMessage class, which
contains the log message, a level such as INFO, WARNING, or ERROR, and a timestamp.
Similar to other objects, it derives from the \SERIALIZABLE class, allowing network transmission in the usual binary encoding.

\subsection{Online Monitoring}
\label{sec:framework:onlinemon}

An important feature of the \EUDAQ framework is the ability to monitor the incoming data stream online using the
\OnlineMonitor, which is provided as a part of the framework.
The \OnlineMonitor only uses the information provided in the byte-stream and does not rely on external information.
It treats every sensor unit as a plane with hits in x-y coordinates and is therefore ideally suited for
pixelated detectors as well as strip-based sensors.
For all planes a standard set of histograms is generated, e.g.\ a hit map both in 1D and 2D and the number of hits per sensor plane.
The \OnlineMonitor also implements some basic clustering routines and the possibility to perform hit correlations between individual sensor planes to enable gauging of the alignment between sensor planes and the particle beam already during data taking.
Additionally, for each different type of sensor, dedicated histograms can be filled if implemented by the user.
Finally, central performance monitoring histograms help to identify potential problems during data taking.

The \EUDAQ \OnlineMonitor uses the \ROOT framework~\cite{Brun:1997pa,Antcheva:2009zz} both for creating histograms and rendering the graphical user interface.
The initial release of the \OnlineMonitor has been described in ~\cite{atlas-troska}, but its performance has been significantly
improved and its functionality extended by the integration of various other devices into the \EUDAQ framework as will be described in Sections~\ref{sec:usage}~and~\ref{sec:application}.

%% file: content/usage.tex
\section{Integration \& Operation}
\label{sec:usage}

The straightforward architecture of \EUDAQ allows new users to quickly integrate existing, hardware-specific DAQ systems into \EUDAQ with minimal overhead.
Within \EUDAQ, each such integrated system is represented by a \Producer.
The \Producer has two roles: to receive commands from \RunControl and to send the acquired data to the \DataCollector.
As such, it provides an interface between the user's DAQ and the rest of the \EUDAQ system.
\Producers can be implemented by deriving from a provided C++ base class, or by using {\tt ctypes}-based Python bindings.
For both use-cases, documentation as well as several examples are provided~\cite{corrin2010, eudaqrepo}.

Whenever a command is received from \RunControl, a corresponding member function of the \Producer is invoked.
By providing their own implementations of these functions, users are able to control their device's DAQ.
The possible commands include \ONINITIALISE, which is called once at startup to perform one-off initialisation;
\ONCONFIGURE, which is called to configure the \Producers, and may be called between runs to reconfigure with different settings;
and \ONSTARTRUN and \ONSTOPRUN, which start or stop the acquisition, respectively.
For the former two, \RunControl includes a list of initialization or configuration parameters and their values, which are retrieved from a plain-text file in order to allow modification of parameters for individual data-taking runs.
The latter two require the \Producers to respond by sending begin-of-run and end-of-run events, respectively, containing specific flags.
As well as ensuring synchronisation of runs between all \Producers, these events also contain the full sensor configuration as well as the user-provided configuration parameters.
Having this information stored alongside the actual data significantly simplifies the bookkeeping for later offline analysis and reference.
A utility program called \MagicLogBook can be used to extract run numbers and their associated parameters from collected data files in order to automatically generate a test beam ``log book''.

During a run, the data retrieved from a device may be submitted through the \Producer's \SendEvent method.
The data is expected to be encapsulated in one of \EUDAQ's event formats:
usually, this is of type \RAWDATAEVENT which is a generic container for blocks of binary data.
Additionally, each event can hold metadata which can be useful to record variables relating to device state changes.
It is up to each \Producer how the raw data is encoded, thus ensuring minimal interference by the data acquisition system
and therefore reducing the potential for data corruption. The data are then serialised and sent to the \DataCollector.

The \DataCollector expects that all devices are synchronised through a hardware signal by a central trigger system and that
their respective \Producers send data on every single event.
Otherwise, the \DataCollector generates an error to alert the user of missing data.

While the data are usually written to disk in the device's native format, for online monitoring purposes and later analysis
stages a conversion is usually necessary.
To easily integrate into the existing tools provided by \EUDAQ, the \DATACONVERTERPLUGIN system was developed.
For each type of \RAWDATAEVENT, a converter plugin can be registered. This plugin will be dynamically loaded and used to
convert the native raw data into a format which can be parsed by \EUDAQ.

Once a \Producer, the \RAWDATAEVENT format and a \DATACONVERTERPLUGIN have been developed, the respective device is fully
integrated into the \EUDAQ system and can seamlessly interact with all
its components.
At this point users are encouraged to submit their code for inclusion into the \EUDAQ software repository~\cite{eudaqrepo}.
This centralised collection of \Producers on the main repository helps prospective users in the detector R\&D community by providing examples. Furthermore, the contributors benefit from the continuous integration methods employed in the \EUDAQ development process.
Through regular automated builds and tests, potential issues can be identified early in the process and mitigated before future releases.
As \EUDAQ is licensed under the GNU LGPLv3~\cite{LGPLv3}, users are generally welcome to share, modify and contribute to its development.
The source code of \EUDAQ is hosted on GitHub, where any potential issues or feature requests can be reported through its issue tracking system.

Since interconnection between different platforms and even different architectures is often necessary when performing beam tests,
\EUDAQ has been developed with interoperability and easy deployment in mind.
Using the \CMAKE build system, \EUDAQ can be compiled on recent versions of GNU/Linux, MacOS~X, and Microsoft Windows using the native compiler tools.
The core library and the command-line executables of \EUDAQ are self-contained and require no additional external dependencies.
Optional components, such as specific \Producers, can be enabled at build-time and might have additional dependencies.
Commonly, these are Qt~\cite{Qt} for the graphical user-interface (GUI) and \ROOT \cite{Brun:1997pa,Antcheva:2009zz} for the online monitoring.

\begin{figure}[htbp]
\begin{center}
\includegraphics[width=0.95\textwidth]{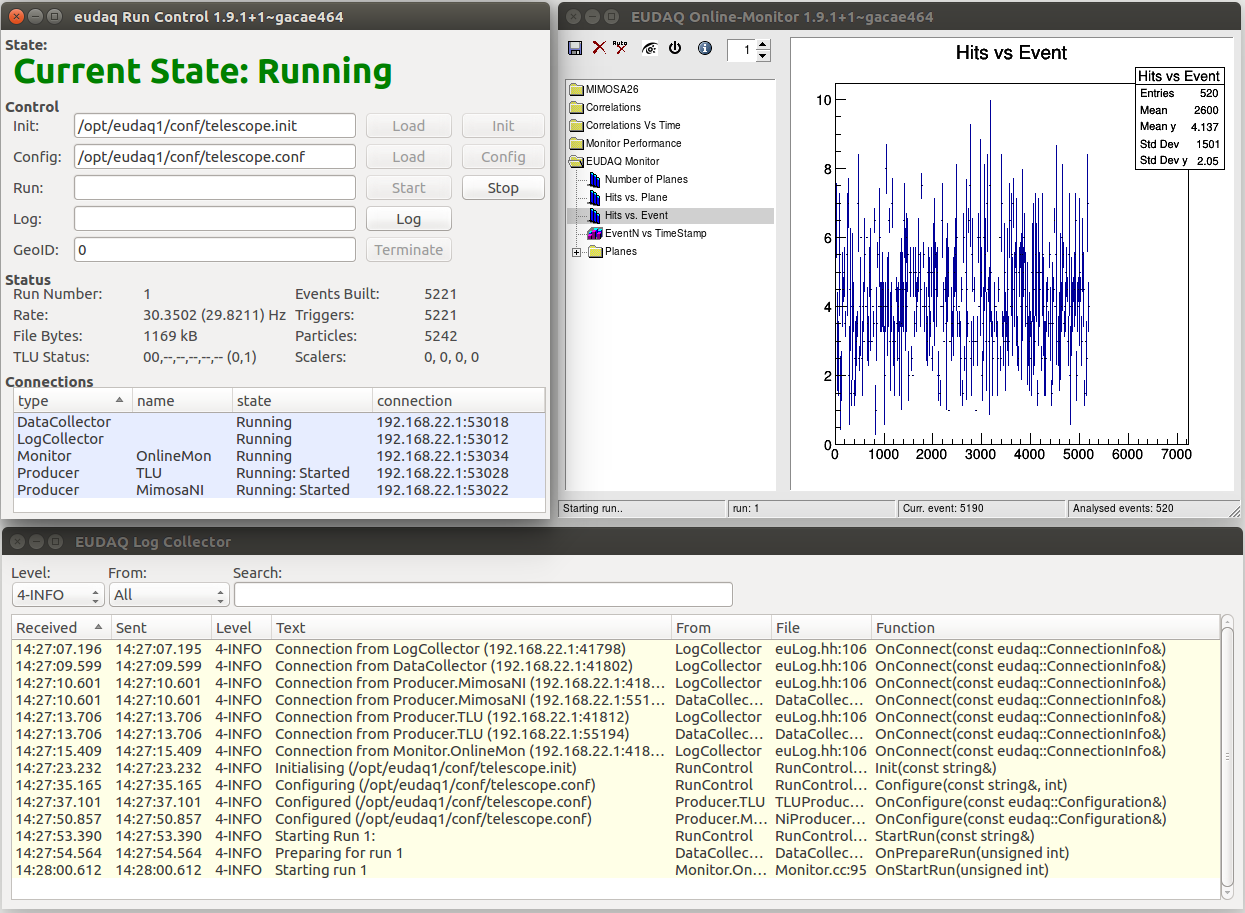}
	\caption{Three graphical user interfaces from \EUDAQ after a standard start-up for running the \EUDET-type telescopes with six \EUDAQ components.
	In the upper left window, the \RunControl shows the global status, the control buttons, detailed status indicators, and a list of all connections. In this example, the \LogCollector, \DataCollector and \OnlineMonitor as well as \Producers for the \MIMOSA sensors and the \EUDET TLU are connected.
	In the upper right window, the GUI of the \OnlineMonitor will show data quality plots and graphs after the data-taking is started.
	In the bottom window, the GUI of the \LogCollector shows messages of the log channel on the selected verbosity level.}
\label{fig:usage:guis}
\end{center}
\end{figure}

Each individual component of \EUDAQ, such as the \RunControl, the respective \Producers and the \DataCollector, run as separate processes
communicating with the rest of the system via TCP/IP on configurable port numbers as described in Section~\ref{sec:framework}.
Start-up scripts are provided that help to set up typical use-case scenarios.
After starting \EUDAQ, three windows are displayed as shown in Figure~\ref{fig:usage:guis}:
firstly, the \RunControl GUI for interaction with the DAQ system and for display of the system state;
secondly, the \LogCollector window for listing all messages reported within the system, allowing direct searches and live filtering;
lastly, the \OnlineMonitor which reads back the data written to disk and generates live-updated histograms.
Together these tools provide the detailed online feedback on the performance of the various components that is crucial to the user during test beam operations.

\EUDAQ is shipped with many additional tools and ready-made \Producers, a complete list of which can be found in the latest manual \cite{eudaqwebsite}.
Some of the optional components are related to the most common use-case: running within the \EUDET-family of beam telescopes.
From a hardware perspective, the beam telescopes consist of six planes of \MIMOSA sensors and a triggering system based around an \EUDET TLU.
Within the \EUDAQ framework, these components are represented by their own \Producers using the approach of integration into \EUDAQ previously described.
The \NIProducer provides the interface to the \MIMOSA DAQ and is built around the National Instrument (NI) PXIe crate architecture \cite{claus2010, claus2011}.
The \TLUProducer connects via USB to the \EUDET TLU to initialise, configure and operate it.
Operating \EUDAQ with the \EUDET-type beam telescopes, the trigger or event rate is limited to 2.0~kHz in average. 
This limit is caused by the data read-out of the \EUDET TLU and the \MIMOSA DAQ, but was also sufficient for most of the \EUDAQ applications in the last decade (see Section~\ref{sec:application}).

For all telescope components, converter plugins are provided as part of \EUDAQ allowing full online monitoring out-of-the-box.
The \EUDAQ converter plugins are also utilised by the beam telescope data analysis framework \EUTELESCOPE \cite{rubinsky2010, bisanz2015, eutel2019}
to interpret the raw data written by \EUDAQ.
\EUTELESCOPE provides the user with extensive tools for clustering, alignment, track reconstruction, and data analysis.
Thus, the telescope hardware with mechanical user interfaces, the \EUDAQ framework, and the reconstruction framework \EUTELESCOPE provide a unique infrastructure, which is extendible for any devices under test and covers all aspects of test beam studies.

To summarise, \EUDAQ plays a central role in a wide-spread set of test beam tools, including the \EUDET-type beam telescopes and the \EUDET TLU:
today there are seven copies of \EUDET-type beam telescopes located at different test beam facilities such as PS and SPS at CERN, the \diitbf, ELSA in Bonn and ESTB at SLAC.
Furthermore, approximately 35 \EUDET TLUs were produced and distributed to different R\&D groups.
Like all infrastructure around the \EUDET-type beam telescopes, \EUDAQ is highly modular and suited for many use-cases as shown in the following chapter.

%% file: content/application.tex
\section{Applications}
\label{sec:application}

EUDAQ is used by a large variety of detector systems.
Many detector prototypes and DAQ systems today provide an integration with the \EUDAQ system and thus greatly simplify their deployment in test beam campaigns.
Especially for users of the \diitbf~\cite{desytb2018}, due to the close integration of the \EUDET-type beam telescopes as well as the
\EUDET TLU described in Section~\ref{sec:usage}, \EUDAQ has proven to be a valuable component of the test beam infrastructure.
In the following, several integrations of detector prototypes with the \EUDAQ system are presented in alphabetical order.
Special attention is paid to the different requirements of the detector systems and the different approaches of integration into the \EUDAQ framework.

\input{content/user/alice}
\input{content/user/atlaspixel}
\input{content/user/atlasstrips}
\input{content/user/belle2}
\input{content/user/calice}
\input{content/user/clicdp}

\input{content/user/hgcal}
\input{content/user/cmspixel}
\input{content/user/cmsp2ot}
\input{content/user/mbi}

\input{content/user/mu3e}

%% file: content/user/alice.tex
\subsection{ALICE Inner Tracker System: Development of MAPS}
\label{sec:application:alice}


The ALICE experiment~\cite{JINST-alice} at the LHC is undergoing a major upgrade in the second Long Shutdown of the collider (2019--2020).
During this upgrade, the Inner Tracking System (ITS) of ALICE will be completely replaced by a new silicon pixel detector.
This new detector consists of seven layers of Monolithic Active Pixel Sensors (MAPS) designed specifically for the upgrade~\cite{alice-tdr}.
This new sensor, called ALPIDE (ALICE PIxel DEtector)~\cite{alice-alpide}, is a $\SI{1.5}{\centi\meter} \times \SI{3}{\centi\meter}$ large pixel detector with 1024~$\times~$512 pixels with a pitch of $\SI{27}{\micro\meter} \times \SI{29}{\micro \meter}$.
The sensor can be thinned down to a total thickness of $\SI{50}{\micro\meter}$.

\begin{figure}[tbp]
  \begin{center}
    \includegraphics[width=0.7\textwidth]{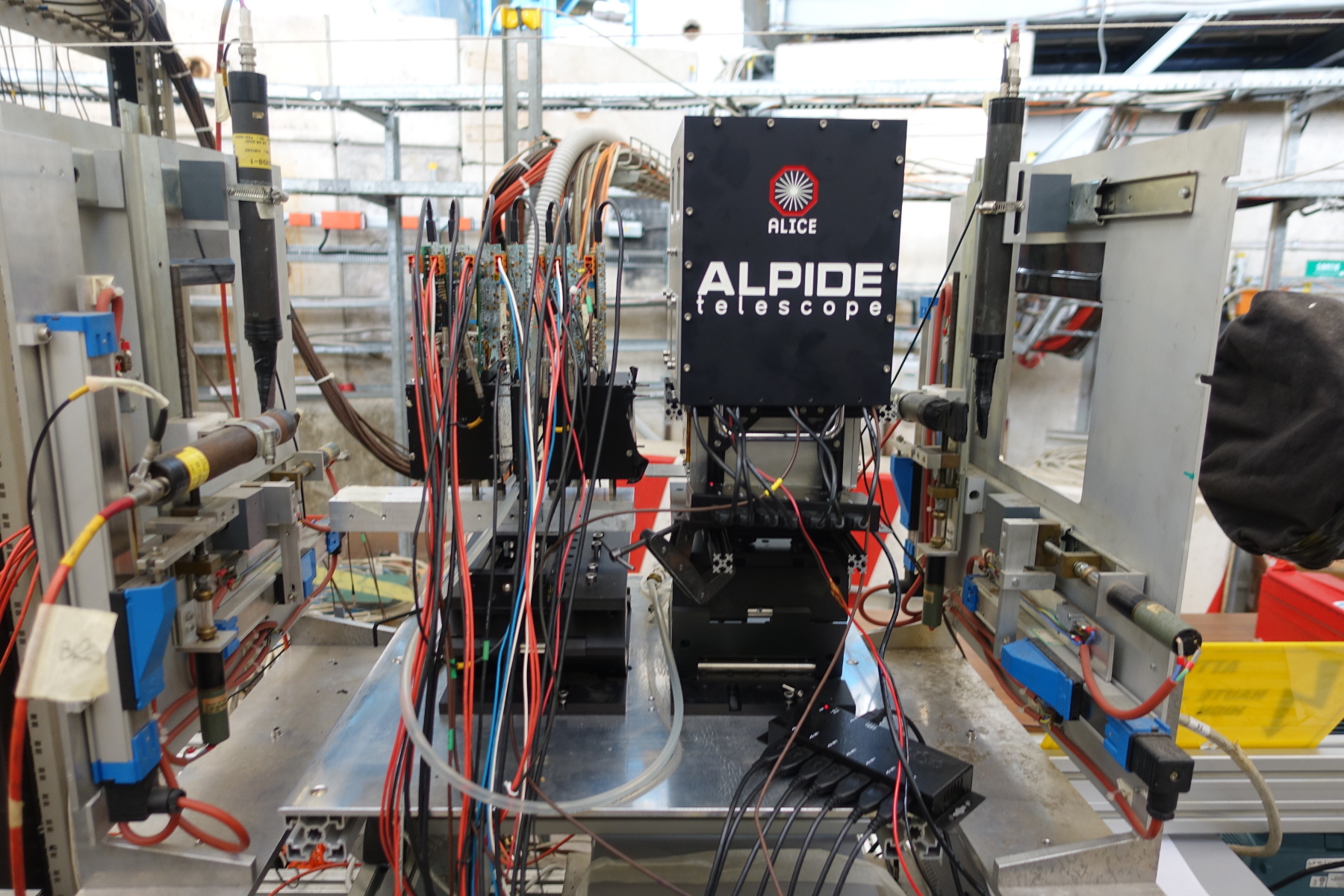}
    \caption{Picture of the self-contained ALPIDE telescope setup, operated and read out using the \EUDAQ framework.}
    \label{fig:ALPIDE_telescope}
  \end{center}
\end{figure}

The ALPIDE sensor has been developed in a long prototyping and testing phase before arriving at the final design.
First prototypes of the detector have been characterised in test beam measurements as early as 2013.
This so-called Explorer chip~\cite{alice-explorer} featured an analogue output, a small active matrix with approximately
12k pixels and has been tested using reference tracks from the \EUDET-type beam telescope.
In a next step, slightly larger prototypes with approximately 32k pixels and an ALPIDE-like architecture~\cite{alice-palpidess}
were tested first with an \EUDET-type beam telescope and later on with ALPIDE prototype-based telescopes as shown in Figure~\ref{fig:ALPIDE_telescope}.
All these different setups have been integrated and were operated using the \EUDAQ framework.

In order to efficiently conduct the many test beam campaigns for the development of the ITS with a total of
approximately \num{20000} runs, a high degree of automation and reliability was required.
This has been achieved by implementing the necessary key features into the \EUDAQ framework:
automatic change of detector configuration at the end of a run; timestamp and data integrity control directly
in the \EUDAQ \Producer; the ability to fully reconfigure the hardware during the run from within the \Producer;
and control of external equipment such as power supplies, rotary and translation stages as well as the cooling unit.

%% file: content/user/atlaspixel.tex
\subsection{ATLAS Inner Tracker: Pixel Detector R\&D}
\label{sec:application:atlaspixel}


The ATLAS pixel detector is the innermost part 
of the tracker subsystem of the ATLAS experiment~\cite{atlas}.
After the initial installation the pixel detector consisted of three barrel layers and 
two end-caps with three disks each. Its modules consist of 
a planar silicon \emph{n\textsuperscript{+}}-in-\emph{n} sensor 
tile to which 16 FE-I3 readout chips are bump bonded. The top of the 
sensor has a flexible PCB attached which hosts an aggregator chip, which multiplexes 
the individual data streams from the ASICS. The default pixel has a 
size of \SI{50 x 400}{\micro\meter} and the detector has achieved a resolution of 
\SI{8}{\micro\meter} in the radial direction ($r\phi$) and \SI{75}{\micro\meter} 
along the $z$-axis~\cite{atlas-pixel-detector}. The BAT telescope~\cite{atlas-bat-telescope} 
based on silicon-strip detectors was used during the R\&D and the qualification phase~\cite{atlas-dobos,atlas-mass} 

In 2010 it was decided to amend the detector with an additional layer of pixel 
detectors, the so-called Insertable B-Layer (IBL)~\cite{atlas-ibl} to improve 
tracking robustness and \emph{b}-quark tagging. At the beginning of the sensor 
oriented research effort FE-I3~\cite{atlas-fei3} based modules were used. Those 
were operated by the laboratory test system TurboDAQ~\cite{atlas-troska}, which 
has been fully integrated into the \EUDAQ framework in order to allow efficient 
testing of the chips using the \EUDET-type beam telescopes. As the development 
of the IBL progressed, both a new readout ASIC, FE-I4~\cite{atlas-fei4}, and a 
new generation of readout systems were introduced and integrated into the \EUDAQ 
framework right from the beginning. In the USBpix Gen2 readout board~\cite{atlas-janssen} the incoming 
data was multiplexed and transmitted by a single \Producer, while with the 
next-generation USBpix Gen3~\cite{atlas-backhaus,atlas-bisanz} the readout was 
fully asynchronous using the so-called STcontrol software~\cite{atlas-stcontrol-buschmann}.
Through this tool, data were copied from the DAQ board to the computer and sent by the 
\Producer simultaneously. A single \COMMANDRECEIVER is instantiated while each connected 
board has its own \DATASENDER which are instantiated once the number of boards 
is known from the configuration. An alternative readout system used for FE-I4 
based modules during test beams is the HSIO-2 system which also provides a 
\Producer.

Different pixel sensor technologies were developed and qualified for the ATLAS IBL in 
multiple test beam campaigns at CERN and DESY: slim-edge  \SI{200}{\micro\meter} thin 
\emph{n\textsuperscript{+}}-in-\emph{n} planar and, 
for the first time-use in high-energy physics, 3D sensors~\cite{PARKER1997328}. The radiation 
hardness of these sensors beyond the required  \SI{5E15}{\neq} was successfully demonstrated. 
The test beam results obtained with the different readout systems and 
sensors~\cite{atlas-ibl-testbeam} may be found in the summary 
paper~\cite{atlas-ibl-modules} and the production 
report~\cite{atlas-ibl-summary}. 

With the IBL installed and operating, the R\&D of ATLAS pixel detector modules 
continued with a focus on the HL-LHC upgrade. A new all-silicon tracking 
detector~\cite{atlas-tdr-phase2} will replace the current inner detector and 
the pixel part will consist of five layers and corresponding end-caps. 
Pixel modules with slim-edge and radiation-hard thin 
\emph{n}-in-\emph{p} planar and 3D sensors were prototyped using the FE-I4 
readout chip and tested up to fluences on the order of \SI{1E16}{\neq} and 
beyond~\cite{Savic_2016,Lange_2018,Bomben_2017}. Two prototypes of readout 
ASICs, FE65-P2 and RD53A~\cite{cern-rd53a}, have been characterised in test beams and for 
the current stage of testing, the RD53A is used.
In order to operate the new generation of readout chips also new readout systems 
YARR~\cite{atlas-yarr, zhang2019measurement} and BDAQ53~\cite{atlas-vogt} have been developed. They 
profited from experience with the previous DAQ generations and both are 
providing an \EUDAQ \Producer. 

Two additional developments originated from within the ATLAS Pixel Detector R\&D:
Firstly pyBAR software for the readout of FE-I4 ASICs has been developed as an alternative DAQ 
software using \python. This has been used with several readout boards including 
USBpix Gen2 and USBpix Gen3 boards. It was initially developed to investigate 
chip tuning methods~\cite{silab-janssen} and then used for hybrid pixel detector R\&D~\cite{silab-pohl2} 
ranging from single-chip readout for sensor characterization~\cite{silab-pohl} to multi-chip readout for 
operation and testing of larger-scale detectors~\cite{silab-gonella, silab-lewis, silab-jaegle}. 
The pyBAR software uses \python bindings to integrate into the \EUDAQ framework 
and to provide a dedicated \Producer which enables the complete control of pyBAR 
via the \EUDAQ \RunControl.

A second development was an additional telescope plane consisting of a FE-I4 based module read out by a 
USBpix board. This has been added to the \EUDET-type beam telescopes to provide a 
reference detector which allows region-of-interest triggering and to provide track 
time stamps with a granularity of \SI{25}{\ns}~\cite{silab-obermann}.

%% file: content/user/atlasstrips.tex
\subsection{ATLAS Inner Tracker: Strip Tracker R\&D}
\label{sec:application:atlasstrips}


For the High-Luminosity LHC the ATLAS experiment~\cite{atlas} will replace the current tracking system with an all silicon detector,
the Inner Tracker (ITk), consisting of inner pixel layers and outer strip layers.
The ITk Strip detector will be composed of four barrels and six end-cap disks on each side, with a total silicon area of approximately 165~m$^2$.
The basic building unit is the strip module, which consists of an $n^+$-in-\emph{p} silicon sensor, one or two hybrids that host the front-end ASICs, 
and a power board, which provides the power for the read-out chips and monitoring functionalities.
The hybrids and the power board are glued directly on top of the silicon sensor~\cite{ATLASStrips}.

Starting from 2014, several prototype modules have been tested during a series of test beam campaigns at DESY, CERN and SLAC,
using \EUDET-type beam telescopes \cite{ITkStripPaper}.
In 2017 the DAQ was upgraded to \EUDAQ version 2 for subsequent test beams \cite{liu2019}.
The readout cycle of the DUTs corresponds to the LHC bunch-crossing frequency at \SI{25}{\nano\second} and is thus much shorter
than the rolling shutter readout of the \MIMOSA sensors of \EUDET-type beam telescope.
In order to select only tracks passing within one readout cycle of the DUT, the integrated FE-I4 plane described in Section~\ref{sec:application:atlaspixel} has been used.
An entire barrel module was irradiated to the full expected dose at the HL-LHC and was characterised in 2016,
which provided valuable information regarding the future performance of the modules at the end-of-lifetime of the HL-LHC.

The ITk Strip DAQ (ITSDAQ) used to operate the detector prototypes runs on an FPGA which performs the trigger-busy-handshake with the \EUDET TLU.
The so-called \ROOT \Producer has been developed to provide the interface between the \EUDAQ configuration file and ITSDAQ~\cite{DissPeschke}.
Two data streams are sent independently to the \EUDAQ \DataCollector: one containing the hit information and timestamps from the
front-end ASICs, and another containing the timing, trigger and control information from the FPGA.
Two converters have been implemented in \EUDAQ for the individual streams.
It has been observed that the hit information data stream sometimes exhibits de-synchronization problems which can, however, be
corrected offline during the analysis of the raw files by comparing the timestamps in the two streams on an event-by-event basis.
To mitigate further problems with de-synchronization, in the latest test beam campaigns an automatic check has been implemented
and deployed in ITSDAQ which allows to directly identify de-synchronised events and re-synchronise the data online.

%% file: content/user/belle2.tex
\subsection{Belle~II Vertex Detector: Development of DEPFET Sensors}
\label{sec:application:belle2}


The Belle~II experiment~\cite{belle2-tdr} is a new general purpose detector operated at the new Japanese Super Flavor Factory, SuperKEKB.
The new machine, working at the intensity frontier, collides electrons against positrons at intermediate energies of \SI{10.36}{\GeV}
but delivering ultra-high luminosities, and is expected to reach \SI{e36}{\per \square \cm \per \second} in the early 2020s.
Belle~II is searching for physics beyond the standard model in B, D and $\tau$ decays through precision measurements
and studies of highly suppressed or forbidden processes~\cite{belle2-physics}.

The main task of the Belle~II vertex detector (VXD) is to provide an accurate measurement of the
impact parameters that allow the precise determination of the position of the vertices of the short-lived particles.
The VXD consists of two layers of DEPFET pixel detectors (PXD) and four layers of double-sided silicon strip sensors (SVD). 
The pixel detector is composed of eight million DEPFET pixels with \SI{50}{\micro\meter} in $r\phi$ and 55-\SI{85}{\micro\meter}
in $z$ and operates at a readout frame rate of \SI{50}{\kHz}.
The reduced thickness of the sensitive volume of \SI{75}{\micro\meter} and the minimal services and support structures sum
up to a total overall material budget of 0.21$\%$~X$_0$.

\begin{figure}[tbp]
	\centering
	\includegraphics[width=0.7\textwidth]{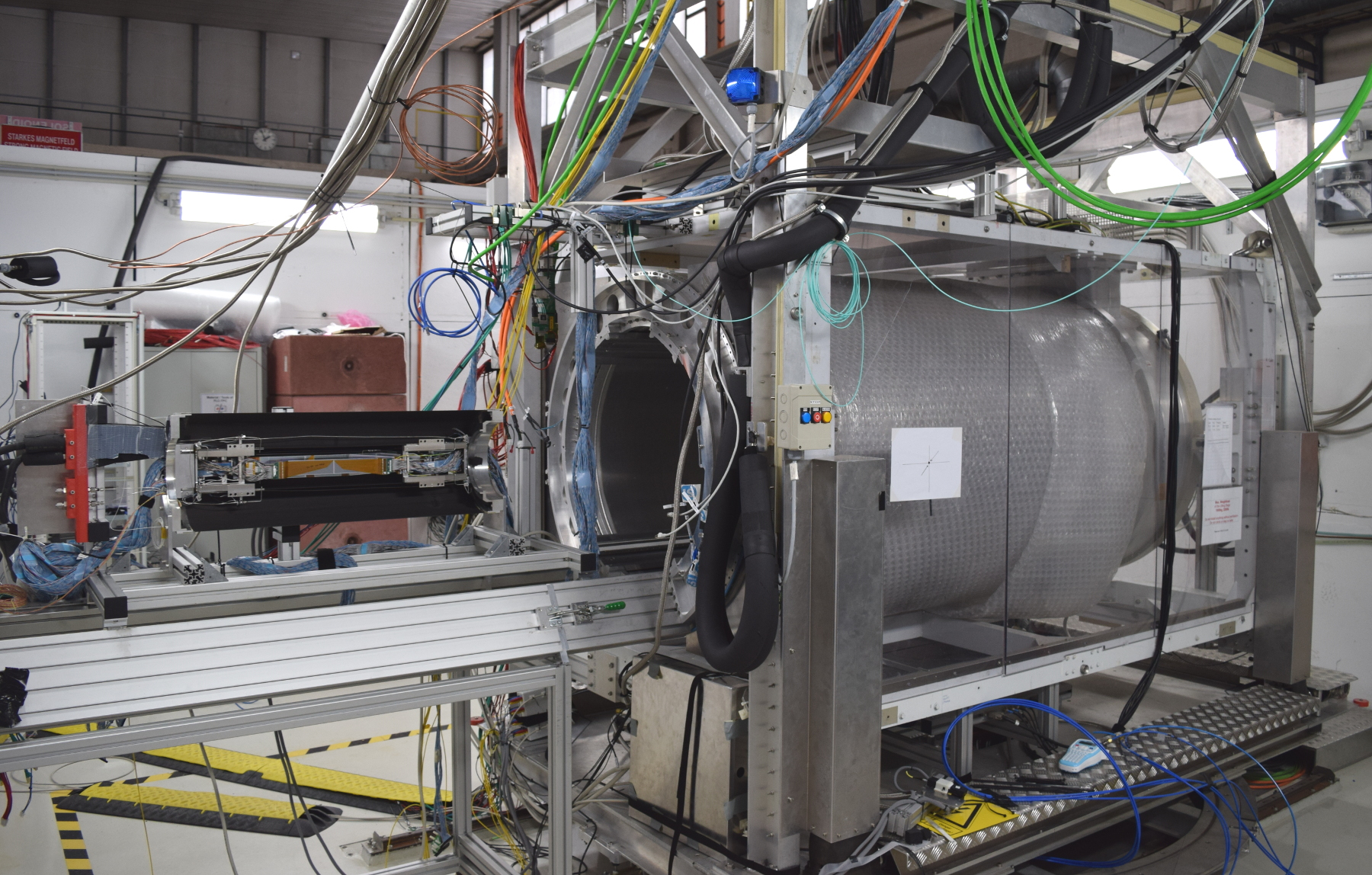}
	\caption{Test beam preparation of the Belle~II Vertex Detector (left) for a full functional test inside a 1~T solenoid (right).}
	\label{belle:tb}
\end{figure}

The final PXD detector modules are the result of more than 10~years of testing and prototyping.
The DEPFET collaboration was the first user to perform a complete integration of the devices under test into the
\EUDET-type pixel telescope~\cite{belle-reuen} including the integration into \EUDAQ.
Early PXD standalone test beams with small scale prototypes focused on measuring the intrinsic spatial resolution and hit
efficiency of \SI{50}{\micro\meter} thick sensors and provided data to validate the detector simulation in the Belle~II software~\cite{belle2-schwenker}.
Several combined VXD and standalone PXD test beam campaigns have been successfully performed~\cite{belle2-bilka, belle2-luck} at the
\diitbf using \EUDET-type telescopes as shown in Figure~\ref{belle:tb}.

In order to increase the data taking efficiency for prototypes with small sensitive area, the DEPFET collaboration used the region-of-interest (ROI) trigger described in Sec.~\ref{sec:application:atlaspixel} to select only tracks traversing the sensitive area of the device under test.
The availability of this FE-I4 reference plane for triggering and timing was valuable also for a test beam with full size (\SI{12.5x44.8}{\mm}) and \SI{75}{\micro\meter} thick DEPFET sensor at DESY in November 2018.
The FE-I4 hit data was read out with pyBAR (c.f.\ Section~\ref{sec:application:atlaspixel}) and sent to the \EUDAQ \DataCollector in order to identify tracks with a resolution of \SI{25}{\ns}
which allowed the measurement of the hit efficiency of the DEPFET devices.

Finally, a main requirement for the Belle~II VXD design was to minimise the material budget as much as possible in
order to reduce multiple scattering of impinging particles.
The magnitude of the effect depends on the local material budget measured in units of the radiation length X$_0$.
The DEPFET collaboration developed a framework to compute high resolution images of the material budget of planar objects
installed in the center of EUDET-type telescopes~\cite{belle2-stolzenberg}, similar to the material budget imaging
described in Section~\ref{sec:application:mbi}.
High resolution images of mechanical prototypes of Belle~II PXD and SVD ladders were obtained in a test beam at DESY in
2015 and were used to validate the \geant model of the Belle~II VXD.

%% file: content/user/calice.tex
\subsection{CALICE: R\&D of Electromagnetic \& Hadronic Calorimeter Prototypes}
\label{sec:application:calice}


The Silicon Tungsten Electromagnetic Calorimeter (SiW-ECAL) and the Analogue Hadronic Calorimeter (AHCAL) are two
CALICE prototypes of high granularity calorimeters for future $e^{+}e^{-}$ colliders.
Both entered the engineering phase of prototyping where the main technological challenges, such as compactness,
reduction of power consumption and embedding of the very-front-end (VFE) in the active layers, are addressed.

The SiW-ECAL technological prototype~\cite{calice-bilokin} consists of thin layers of silicon sensors (300-\SI{650}{\micro\meter})
as active material alternated with tungsten plates as absorber material.
The active layers are made of a matrix of \si{2x2} silicon sensors segmented each in matrices of \si{8x8} squared pixels with a pitch of \SI{5.5}{\mm}.
The VFE consists of 16~SKIROC ASICs~\cite{calice-callier} that have been designed
for the readout of silicon PIN diodes.
The AHCAL prototype consists of detector layers that are made from \SI{36x36}{\cm} units, each equipped with
144 scintillator tiles ($30 \times 30 \times \SI{3}{\cubic \mm}$) individually read out by SiPMs.
The VFE consists of four SPIROC ASICs~\cite{calice-conforti}.

The prototypes use different DAQ systems.
While the AHCAL readout is based on a Labview DAQ, integrated into the \EUDAQ framework by means of a \Producer,
the independent Calicoes software~\cite{calice-magniette}, based on the Pyrame framework~\cite{calice-rubio}, has been
developed for control and data acquisition of the SiW-ECAL.
Both prototypes are self-triggered and are designed for the ILC~\cite{ilc-tdr} timing structure with bunch trains of
high activity interspersed with significant periods without beam which are used for readout and lead to dead time.
Therefore the synchronization of the active periods of the detectors is very important for test beams.

The hardware synchronisation was ensured by two clocks -- \SI{40}{\MHz} and \SI{50}{\MHz}, derived from a single clock
generator -- and a shared signal to control the acquisition.
The \EUDAQ framework was used to integrate the two separate DAQ frameworks of the prototypes.
\EUDAQ producers receive data from each subsystem via a TCP/IP socket connection
and send them to the \DataCollector to combine them into one event in \LCIO format by correlating their timestamps.
\LCIO collections are used to separate the individual subsystems.

In 2016, test beams of the AHCAL prototype with an externally-triggered \EUDET-type beam telescope were conducted at the \diitbf.
Hardware synchronization was provided by an \EUDET TLU running in the trigger-handshake mode.
The trigger signals were recorded by the AHCAL DAQ hardware to allow correlation with the self-triggered hits from the AHCAL.
Events were then combined on the basis of the trigger number and converted to \LCIO.
The \EUDAQ online monitor provided a visual verification of the synchronization via the spatial correlation between the detectors.

\EUDAQ was also used for a combined test at the CERN~SPS in 2017 of the AHCAL prototype with the CMS~HGCAL prototype
described in Section~\ref{sec:application:hgcal}.
For this, an additional \EUDAQ \Producer was developed to control and read out Delay Wire Chambers (DWCs) using a CAEN TDC VME module.
Similar to the preceding test beams, events were built in \EUDAQ using the trigger numbers recorded by all participating devices.

Since the first test beam with \EUDAQ, the AHCAL group has adopted \EUDAQ as DAQ framework for most
test beams including the integration with the \AIDAII beam telescopes and the TLU.
Multiple tests were performed in different test beam campaigns, which were crucial in the development of \EUDAQII,
supported by the \AIDAII project work package for common test beam DAQ developments.
The CALICE collaboration has decided to encourage all participating groups to use \EUDAQ for new common test beams~\cite{calice-wing}.

%% file: content/user/clicdp.tex
\subsection{CLIC Vertex \& Tracking Detectors: Pixel Detector R\&D}
\label{sec:application:clicdp}

The CLICdp collaboration studies available silicon pixel detector technologies,
evaluates their tracking performance and identifies promising candidates for the
construction of CLIC tracking and vertex detectors. Low material budget, small
pixel pitch for a good spatial resolution and a few-nanoseconds timing
resolution are needed to satisfy the physics requirements and cope with the
experimental conditions in the CLIC detector.

In the context of this study, the Timepix readout ASIC \cite{clicdp-timepix} was used as a test vehicle to study
the tracking characteristics of pixel sensors of various thicknesses and different polarities in order to develop
simulation models to estimate the performances of future prototypes with smaller pitch and thinner substrate as
required for CLIC.
A series of sensors with thickness varying from $\SI{50}{\micro\meter}$ to $\SI{500}{\micro\meter}$, with both
polarities, was produced and a systematic characterization of their tracking properties using the \EUDET-type telescope,
the \EUDAQ software and custom \EUDAQ~\Producers was performed.

\begin{figure}[tbp]
  \center \includegraphics[width=0.4\textwidth]{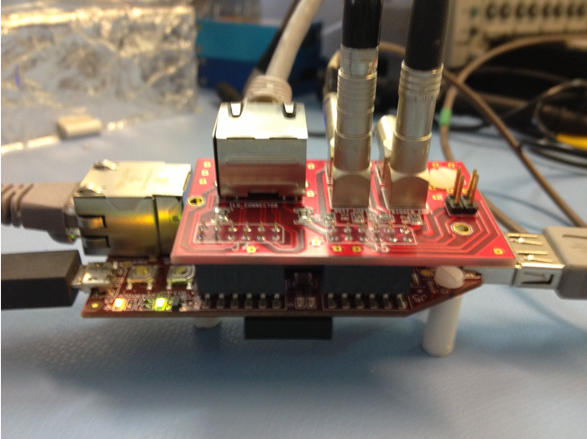}
  \caption{The MIM-TLU hardware board}
  \label{mimtlupic}
\end{figure}

For initial studies, the proprietary readout system FITPix \cite{clicdp-fitpix} was used.
The pre-compiled libraries as well as the headers were provided by the Timepix/Medipix collaboration to allow
interfacing the system with \EUDAQ, but no modifications of firmware or software were possible.
With the frame length of the \EUDET-type beam telescopes of up to \SI{230.4}{\us} being much smaller than the configured
frame duration of the Timepix, a simple trigger-busy scheme providing only one track per trigger issued would have severely limited the data rate.
To mitigate this, an intermediate hardware interface between the TLU and the FITPix readout hardware, The Man-In-the-Middle
TLU Unit (MIM-TLU) show in Figure~\ref{mimtlupic}, was designed.
It consists of a Spartan6 FPGA evaluation board and a custom shield board providing the interface to the RJ45 DUT
connection of the TLU and the Trigger-Busy interface of the FITPix.
It allows the duty cycle of the Timepix detector to be maximised by accepting a configurable number of triggers from
the TLU for each frame of the Timepix.
The events were synchronised based on trigger IDs and the duty cycle of the system could be improved from \SI{1.13}{\percent} to \SI{40}{\percent} via the integration of up to 100 TLU triggers.

The Timepix \EUDAQ \Producer also provides interfaces to a Keithley High-Voltage power supply through a GPIB-USB
interface and to a rotation stage via RS232 in order to perform automated threshold, bias and incident angle scans.
The scan parameters were provided to the producer through the standard \EUDAQ configuration files and the
information on the system configuration was encoded in the header of the \EUDAQ data files for further reference.
The \MagicLogBook tool was used to extract the conditions for each run and automatically populate the test beam log with the appropriate information.
Finally, a decoder for online monitoring and integration to the \EUTELESCOPE reconstruction framework was developed in
order to verify data quality during test beam and to perform reconstruction of the events for further analysis.
More than nine weeks of continuous test beam was performed with this setup and the substantial amount of data
acquired lead to multiple publications and PhD theses \cite{clicdp-timepix-calib,clicdp-timepix-calib-results,clicdp-alipour,clicdp-ccpdv3,clicdp-buckland}.

Following the successful experience with the Timepix integration, a Timepix3 producer was created and the FITPix system was replaced by the SPIDR readout \cite{clicdp-spidr}.
The Timepix3 uses a data-driven acquisition model with no dead time for the readout and no additional hardware was required for operation with the \EUDET-type telescope.
Similarly, a \Producer for the CLICPix ASIC \cite{clicdp-clicpix} was produced, based on the $\mu$ASIC system.
Results of the Timepix3 and CLICPix characterisation can be found elsewhere \cite{clicdp-technologies}.

%% file: content/user/hgcal.tex
\subsection{CMS High Granularity Calorimeter for HL-LHC}
\label{sec:application:hgcal}


For High-Luminosity LHC the CMS experiment will replace its existing end-cap (1.5<$|\eta|$<3) calorimeters
 with a sampling calorimeter based on a mixture of silicon sensors in the highest radiation regions, and scintillating tiles with on-tile SiPM readout.
The resulting High Granularity Calorimeter (HGCAL) will have unprecedented readout and trigger granularity for particle showers,
facilitating particle-flow analyses to ensure good energy resolution and particle identification in the
very high pileup and radiation environment expected \cite{cms-tdr-phase2-hgcal}.
The HGCAL will be installed during Long Shutdown 3 of the LHC, foreseen around 2024-2026.
A series of beam-tests was planned in 2015, in order to validate the feasibility of the technically-challenging HGCAL
design and to compare its simulated performance with that found in reality.

For the first small-scale tests of prototype hexagonal silicon modules at FNAL and CERN in
2016 \cite{akchurin2018} a proprietary DAQ system \cite{rubinov2016} was sufficient.
Larger-scale tests were planned for 2017/18 using new DAQ components.
In order to allow synchronous data taking with other hardware, \EUDAQ was selected as central DAQ
system due to its configurability, scalability, integrated Data Quality Monitoring (DQM), ease of use and relatively simple integration with different hardware.
Initially three dedicated \EUDAQ \Producers were implemented: for the so-called RDOUT boards using
the IPbus protocol~\cite{larrea2019}, for CAEN v1290 TDCs, and for CAEN v1742 digitisers.
The latter were incorporated for the readout of various beam-characterizing detectors at CERN
such as Delay Wire Chambers (DWCs) \cite{wirechamber-guide}, threshold Cerenkov counters and timing devices.

Following the recommended procedure, \DATACONVERTERPLUGINS for the online reconstruction and
analysis of the HGCAL data were developed.
The converted data were then visualised by suitably modifying the \EUDAQ DQM to correctly display the hexagonal pixels of the CMS HGCAL modules.
Independently-developed systems using \EUDAQ for the readout, such as the CALICE AHCAL presented
in Section~\ref{sec:application:calice}, could be integrated into the tests with little modifications to the triggering logic at the hardware level.
It is noteworthy that for the 2018 HGCAL test at DESY, the integration of the \EUDET-type telescopes
and X-Y stage was straightforward and allowed exploiting the pre-installed telescope \Producer, \SlowProducer and the TLU provided at the \diitbf.

\begin{figure}[tbp]
  \begin{center}
    \includegraphics[width=0.65\textwidth]{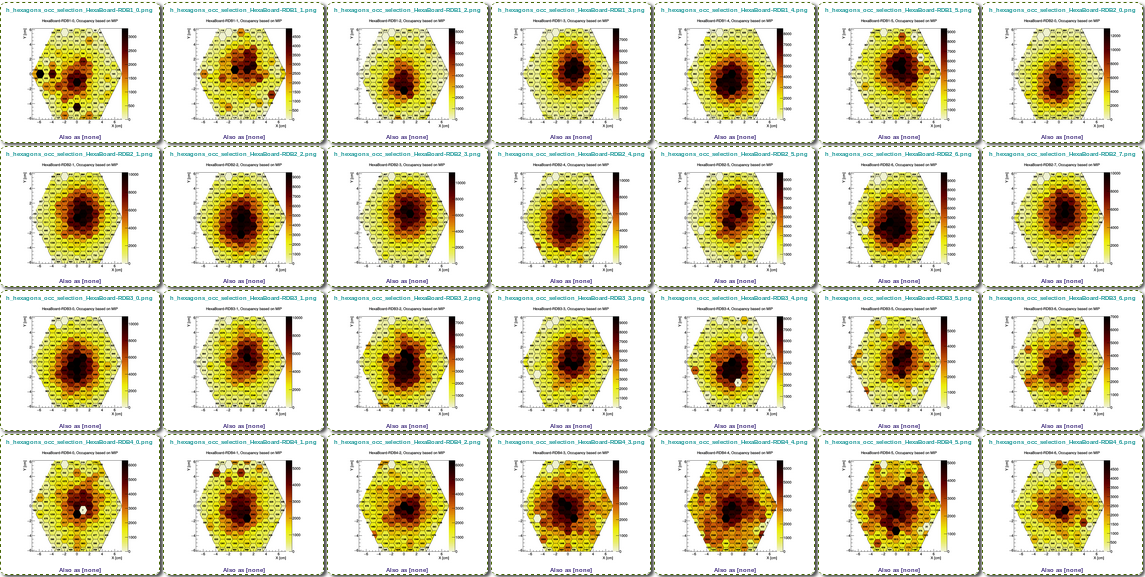}
	  \caption{Screenshot of the \EUDAQ DQM, here showing the pixel occupancy in 28 single-module planes tested in 2018, due to incident
	  100~GeV electrons, showing good centering of the beam and the development of the electromagnetic shower.
	  Note that the orientation of the planes in the DQM do not always represent reality.}
    \label{fig:hgcal_dqm}
  \end{center}
\end{figure}

Using \EUDAQ as central DAQ framework, systems ranging from a single 128-channel hexagonal silicon module,
to 94 modules plus the 39-layer 2018 CALICE AHCAL prototype, were successfully tested in beams at DESY and CERN in 2017 and 2018 for a total of more than two months.
The final test at CERN in October 2018 was performed with a prototype comprising a total of \num{34000} channels.
The \EUDAQ-based DAQ was reliable and sufficiently simple to be operated by 40 trained shifters in total.
Data from up to four \EUDAQ producers, the three mentioned previously plus one for the AHCAL, were acquired
at a rate of around \SI{50}{\Hz}, limited by the DAQ hardware.
In total, more than 6~million pion, electron and muon events were recorded for analysis.
This in-built DQM proved very useful for ensuring good data quality, as can be seen in Figure \ref{fig:hgcal_dqm}.
Data are currently being analyzed, but initial indications are very positive.

%% file: content/user/cmspixel.tex
\subsection{CMS Pixel Detector: Test Beams for the Phase I Upgrade}
\label{sec:application:cmspixel}

In the year-end technical stop 2016/2017, the CMS experiment has replaced its pixel detector with a new detector system
in order to withstand the higher instantaneous luminosity of the LHC of $\mathcal{L} = 2\times10^{34}\,\mathrm{cm^{-2}s^{-1}}$
expected for the following years of operation~\cite{cms-tdr-phase1}.
This so-called Phase-I Upgrade comprised, among other changes and improvements, of a new front-end ASIC, also referred to as readout chip (ROC).
This chip is an evolution of the ROC used in the first pixel detector for CMS, implementing several enhancements
to allow for higher particle rates and occupancy.
The chip is fabricated in a radiation-hard $\SI{250}{\nano \meter}$ process and features a pixel pitch of
$\SI{100}{\micro \meter} \times \SI{150}{\micro \meter}$ and a matrix of $80 \times 52$ pixels.

\begin{figure}[tbp]
  \centering
  \includegraphics[width=.7\textwidth]{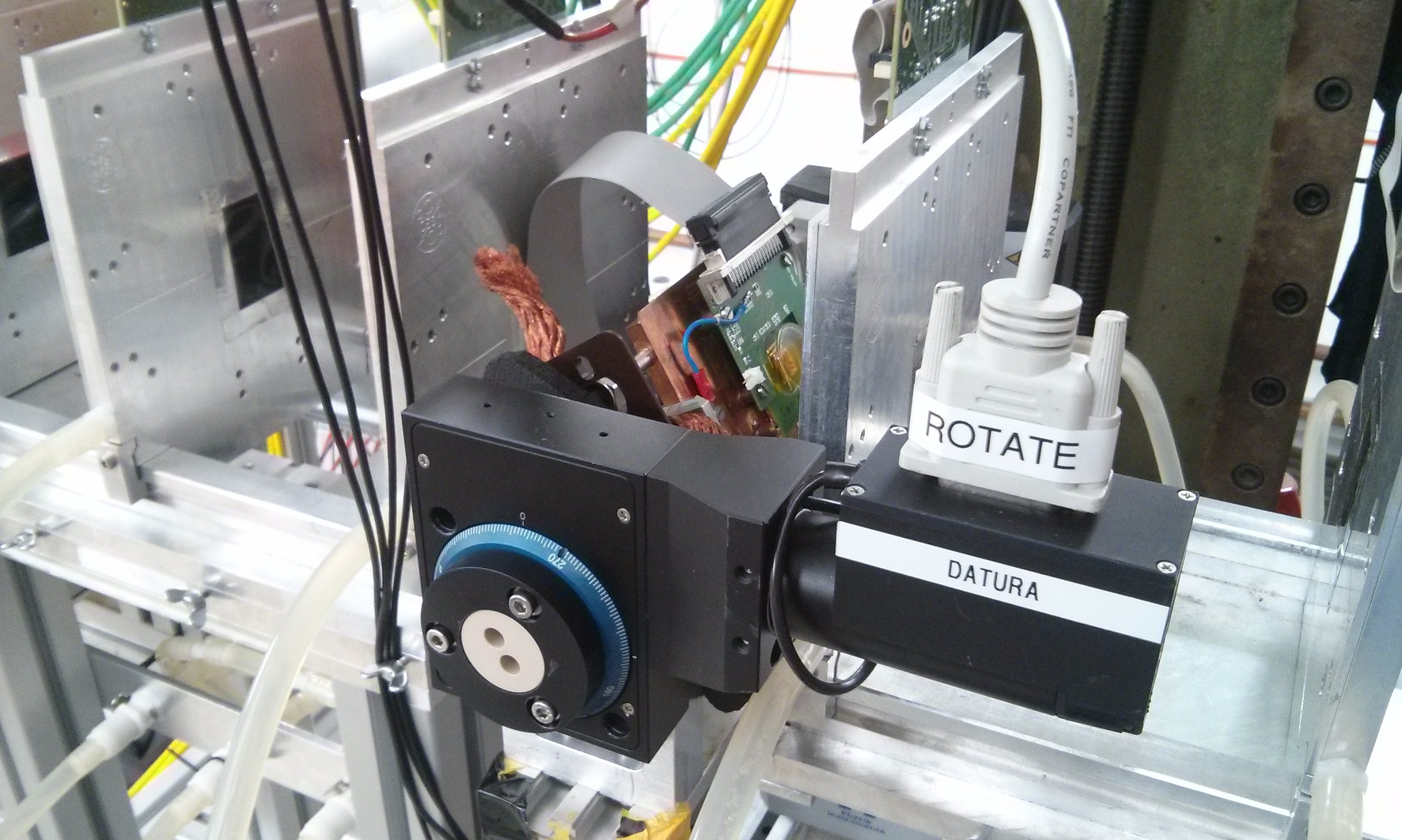}
  \caption{The CMS pixel detector ROC mounted between the two arms of a EUDET-type beam telescope. From~\cite{cmspixel-spannagel}.}
\label{fig:cmsp1:newdaq}
\end{figure}

Many test beam campaigns have been conducted in order to test new ROC versions, to verify design changes
and to ensure good performance of the final detector modules~\cite{cmspixel-testbeams, cmspixel-testbeam-paper}.
These measurements have been conducted at the \diitbf using positrons from the \desyii synchrotron.
Assemblies with the different ROC versions have been operated as DUTs and were placed
between the two arms of an \EUDET-type pixel beam telescope as shown in Figure~\ref{fig:cmsp1:newdaq}.
In addition to the DUT in the center of the beam telescope, a second ROC was placed downstream at the end
of the beam telescope as a timing reference.
This is necessary for efficiency measurements due to the rolling-shutter readout of the telescope detectors
with an integration time of $\mathcal{O}(\SI{100}{\micro \second})$ and the resulting track multiplicity in the beam telescope.

A dedicated readout system has been designed and built for the CMS pixel detector ROC family, consisting of a
custom-designed FPGA-based readout board and the software library \emph{pxarCore}~\cite{cmspixel-pxar}.
The system is capable of operating single ROC assemblies as well as full detector modules and has been used for
test beam measurements and for the qualification of all final detector modules produced.

The software library strictly separates user inputs from hardware calls by means of a hardware abstraction layer (HAL).
This concept has allowed a stable application programming interface (API) to be provided to various users and developers
of user interfaces while major changes could be implemented on firmware and HAL-level without entailing additional changes
to user software or breaking compatibility with older versions.
Using this interface, an \EUDAQ \Producer (CMSPixel) was written which parses the configuration parameters obtained
from the \EUDAQ \RunControl and calls the appropriate API functions of the \emph{pxarCore} library.
Providing a \DATACONVERTERPLUGIN for \EUDAQ ensured full integration of the detector into the online monitoring
system and allowed supervision of basic detector parameters as well as correlation with telescope planes
during data taking. 

The test beam activities and results have been summarised in a publication~\cite{cmspixel-testbeam-paper}, and the
detector was successfully built, commissioned and is in operation since the LHC year-end technical stop 2016/2017.
The test beam setup and its \EUDAQ integration has also been used for measurements studying the performance of the
ROC under the influence of radiation-induced damage~\cite{cmspixel-schuetze}.

In addition to the comprehensive test beam campaigns for the characterization of the new readout chip, measurements
with full CMS Phase~I pixel detector modules have been performed~\cite{cmspixel-schnake}.
Here, a telescope consisting of four detector modules with 16~ROCs each has been operated using the same readout electronics.
In order to correlate the data from the different planes and to orchestrate the start and stop of a run, the \EUDAQ \RunControl
was used together with four individual \emph{CMSPixel} producers, one for each telescope plane.
Owing to the flexibility of the \EUDAQ framework and the underlying DAQ library, this did not require any changes to the code but provided an out-of-the-box experience for data acquisition of the new telescope, including online monitoring capabilities.

%% file: content/user/cmsp2ot.tex
\subsection{CMS Outer Tracker: Module Qualification in Test Beams for Phase~2}
\label{sec:application:cms2otk}

\begin{figure}[tbp]
    \centering
    \includegraphics[width=.7\textwidth]{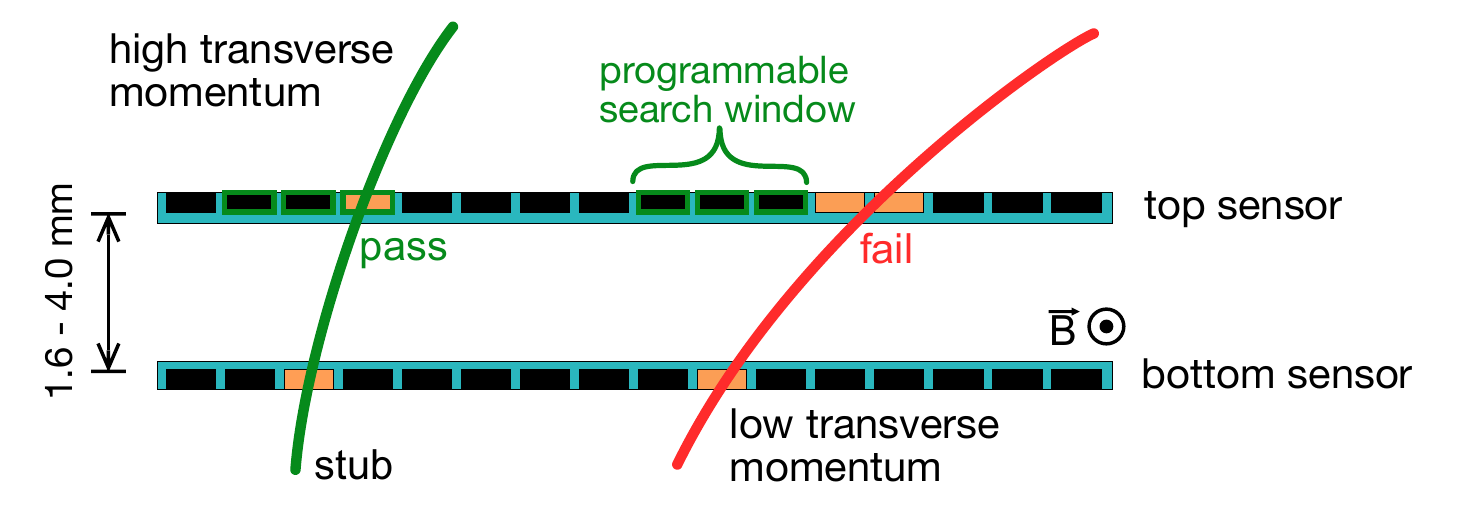}
    \caption{\label{cmsp2ot:stub} Example of the on-module $p_{T}$ discrimination.
    The left track has a higher transverse momentum, thus registering a hit on the upper sensor within the programmable search window.
    The low-momentum right track fails this cut.}
\end{figure}

For the HL-LHC, the CMS experiment will substantially upgrade its tracking detector~\cite{cms-tdr-phase2}.
The all-new CMS Phase-2 tracker will consist of an Inner Tracker (IT) based on silicon pixel modules and an Outer Tracker
(OT) made from silicon modules with strip and macro-pixel sensors.
In the Outer Tracker, so-called $p_{\rm T}$ modules will be utilised, which will provide input from the OT to the CMS L1 trigger.
The $p_{\rm T}$ modules are composed of two closely-spaced silicon sensors, which are read out by a common front-end ASIC,
correlating the signals in the two sensors.
With the $3.8$~T field of the CMS magnet, tracks from charged particles are bent, with the helix radius depending on
the $p_{\rm T}$ of the particle, as shown in Figure~\ref{cmsp2ot:stub}.
Hit pairs corresponding to signals compatible with particles above the chosen $p_{\rm T}$ threshold are called stubs and are sent to the L1.

Two different module types are foreseen for the OT: 2S modules comprising two strip sensors, and PS modules,
where a strip sensor and a macro-pixel sensor are used.
The latter are deployed at radii from $200-600$~mm, the former at radii larger than $600$~mm.

Four successful test beam campaigns investigating Outer Tracker prototype modules have been performed at DESY and
CERN in 2017 and 2018 with the \EUDET-type beam telescopes.
The FPGA running the CMS OT DAQ performs the full trigger-busy-handshake protocol with the \EUDET TLU.
A custom \EUDAQ \Producer has been developed and interfaces the CMS OT DAQ with the \EUDAQ \RunControl.
Whilst the initial configuration of the CMS modules is performed separately, the \Producer sends the hit data of
both prototype sensors to the \DataCollector.
Furthermore, parameters from the front-end ASICs such as TDC or stub information are added to the individual
\EUDAQ events for the track reconstruction and subsequent analysis.

%% file: content/user/mbi.tex
\subsection{Material Budget Imaging: Using Scattering Angles to Gauge Material}
\label{sec:application:mbi}

\begin{figure}[tbp]
	\centering
	   \begin{subfigure}[b]{0.8\textwidth}
	      \centering
	      \caption{}
	      \includegraphics[width=\textwidth]{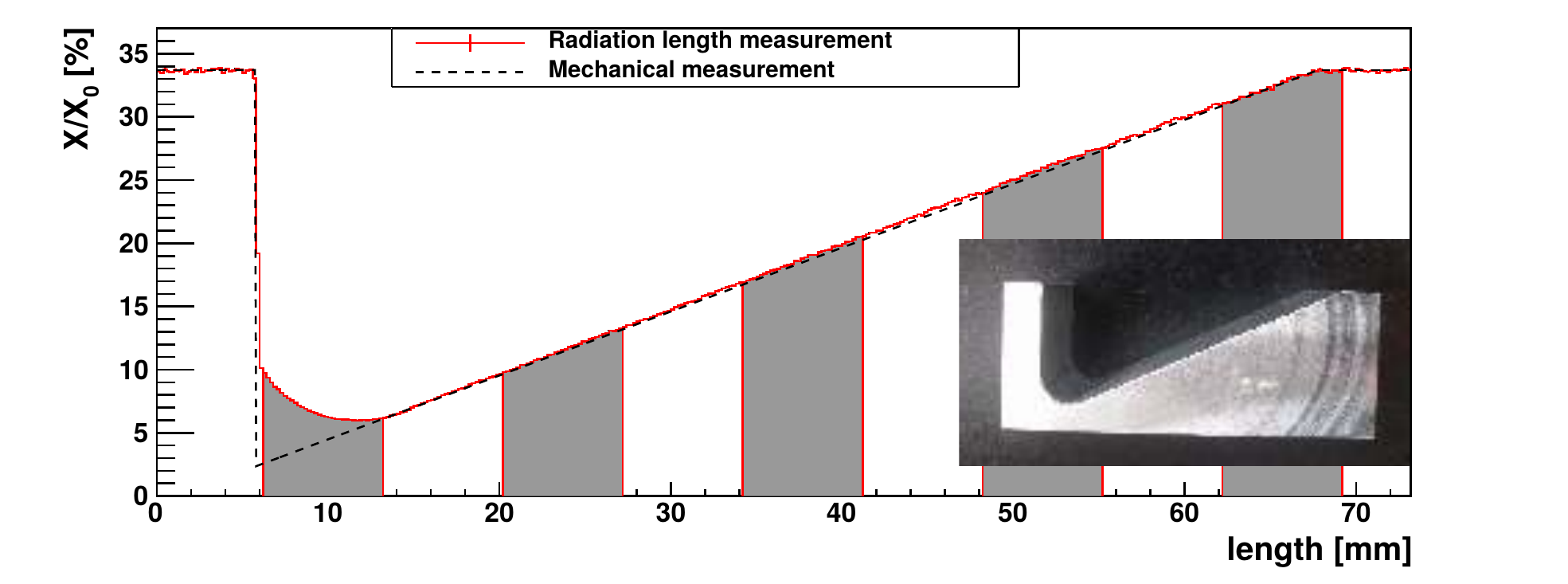}
		\label{fig:mbi_examples:a}
	   \end{subfigure}
	   \qquad
	   \begin{subfigure}[b]{0.75\textwidth}
	      \centering
	      \caption{}
	      \includegraphics[width=\textwidth]{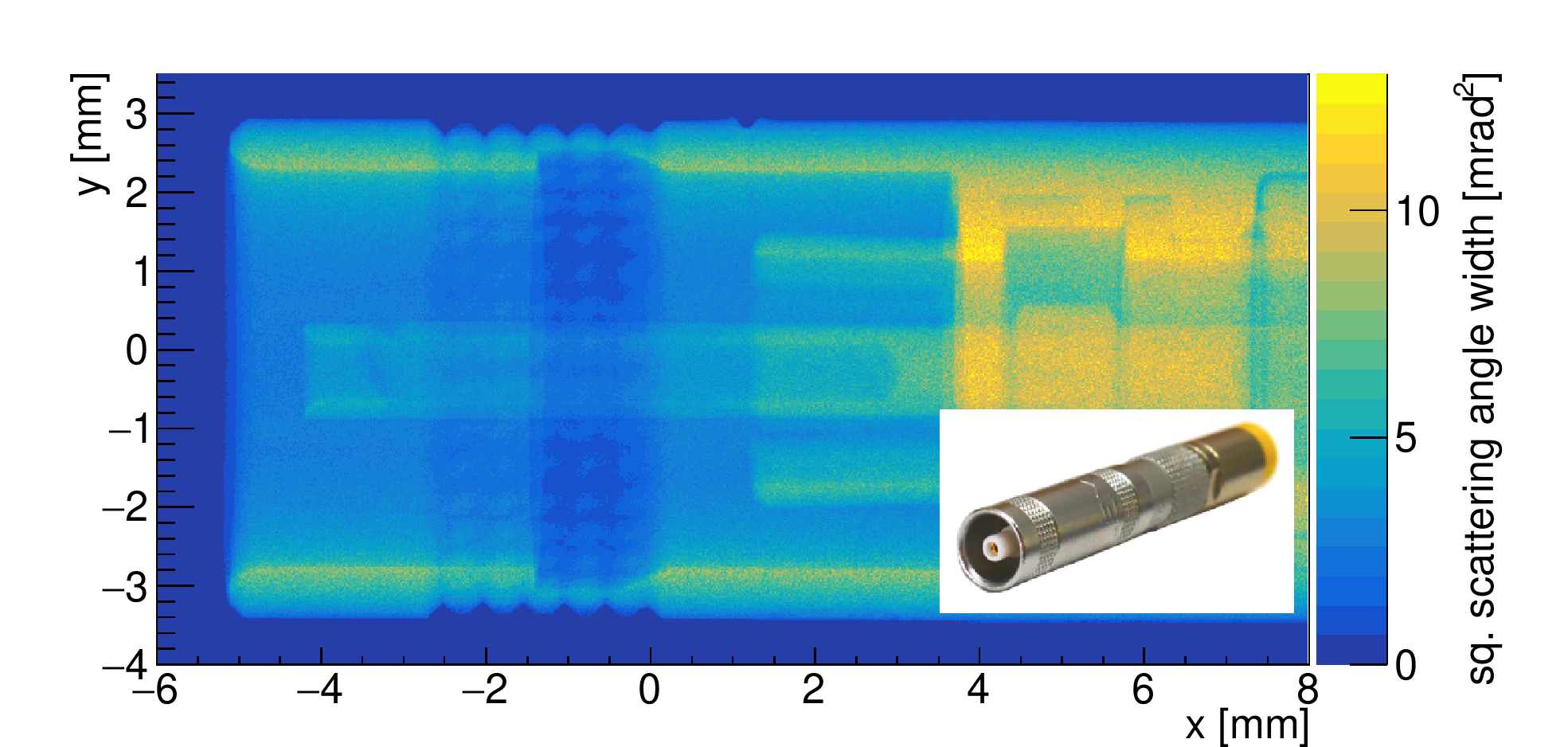}
		\label{fig:mbi_examples:b}
	   \end{subfigure}
	\caption{Material budget imaging: 
	(a) Calibration scan using an aluminium wedge with values ranging from 6.7~\% (6~mm aluminium) to 33.7~\% (30~mm aluminium) including a statistical uncertainty of approximately 0.4~\%. The alternating gray and white regions indicate different measurement positions on the wedge. The measurements are based on a total sample of 470 million tracks. 
	(b) Two-dimensional scan of a LEMO connector resolving the inner substructure.}
	\label{fig:mbi_examples}
\end{figure}

The material budget is the ratio of the thickness of a sample and its radiation length X$_0$.
One of the main concerns when designing tracking detectors for particle detectors is to minimise this value in
order to reduce the effect of multiple scattering which degrades the achievable track resolution of the detector.
Combining the \EUDET-type telescopes as high-precision tracking devices and the
moderate test beam momenta of the \diitbf, 
a high spatial resolution for material budget imaging measuring the scattering angle of each particle can be achieved.
This methodology has been used by different groups and experiments to gauge the material budget of detector prototypes or other objects, for example~\cite{belle2-stolzenberg, thesis-stolzenberg, jansen2018}.

Using this method, structures of devices can be resolved in a two-dimensional image covering the \MIMOSA
active area or even in three-dimensional tomographic images by rotating the sample under test~\cite{jansen2018tomo, cmspixel-schuetze}.
In this case a large amount of recorded particle tracks is needed to reach a reasonable resolution and contrast in the images.
With the particle fluxes in the order of \SI{5}{\kHz\per\m^2} achievable at the \diitbf, it is required to employ automatic data taking over longer time periods.
\EUDAQ and the DAQ of the \EUDET-type beam telescopes have shown to perform stably when conducting measurement campaigns overnight using the automatic-next-run option.

The $x$ and $y$ linear stages can be used in order to explore samples larger than the active
area of the \MIMOSA sensors.
The rotational stage offers the possibility to perform tomographic studies.
For this, the control over the different motor stages was implemented in \EUDAQ as a \SlowProducer, which allows
one-dimensional scans to be performed by defining the scan range in the configuration file.
For performing arbitrary two-dimensional scans, multiple configuration files for the respective coordinates can be created.
After the completion of a given run, corresponding to one configuration file, the subsequent file is automatically loaded by the \RunControl.
This feature is extensively used for the three-dimensional tomographic imaging technique.

Owing to this fully automated data taking as well as the possibility to record even larger areas than the size of the \MIMOSA active
area by stitching images, this technique is a valuable tool to gain insights into the material budget of calibration material as well as complex
structures as demonstrated in Figure~\ref{fig:mbi_examples}.

%% file: content/user/mu3e.tex
\subsection{Mu3e: Development of HV-MAPS Sensors}
\label{sec:application:mu3e}

The Mu3e experiment will search for the charged lepton flavor violating decay $\mu^+\rightarrow e^+\,e^-\,e^+$.
The low momenta of the decay particles requires an ultra thin pixel sensor with high rate capabilities. Additionally,
the continuously decaying muons do not allow for a hardware trigger, demanding a zero suppressed streaming readout
of the pixel sensors. High-voltage monolithic active pixel sensors (HV-MAPS)~\cite{mu3e-peric} are chosen for the
Mu3e tracker, as they offer a high integration level combined with the option of thinning the sensor down to a
total thickness of only \SI{50}{\micro\meter}.

Four HV-MAPS prototype generations have been studied utilizing the \EUDET-type telescopes during 14 test beam campaigns at the \diitbf.
The triggerless fast sensor readout poses a challenge for the integration into the triggered \EUDAQ framework.
A hit is registered and read out in less than \SI{200}{\nano\second}, compared to the rolling shutter revolution
time of up to \SI{230.4}{\micro\second} of the \MIMOSA sensors.
During a single \MIMOSA readout cycle, the timestamps on the MuPix laps 15~times.
Additionally, the prototypes are significantly smaller than the \MIMOSA sensors, leading to events without a hit in the MuPix.
To avoid several triggers without hit-data, empty blocks are sent to the \DataCollector to match the trigger frequency.
The full trigger-handshake with the TLU is realised on an FPGA.

\begin{figure}[tbp]
	\centering
	\includegraphics[width=.6\textwidth]{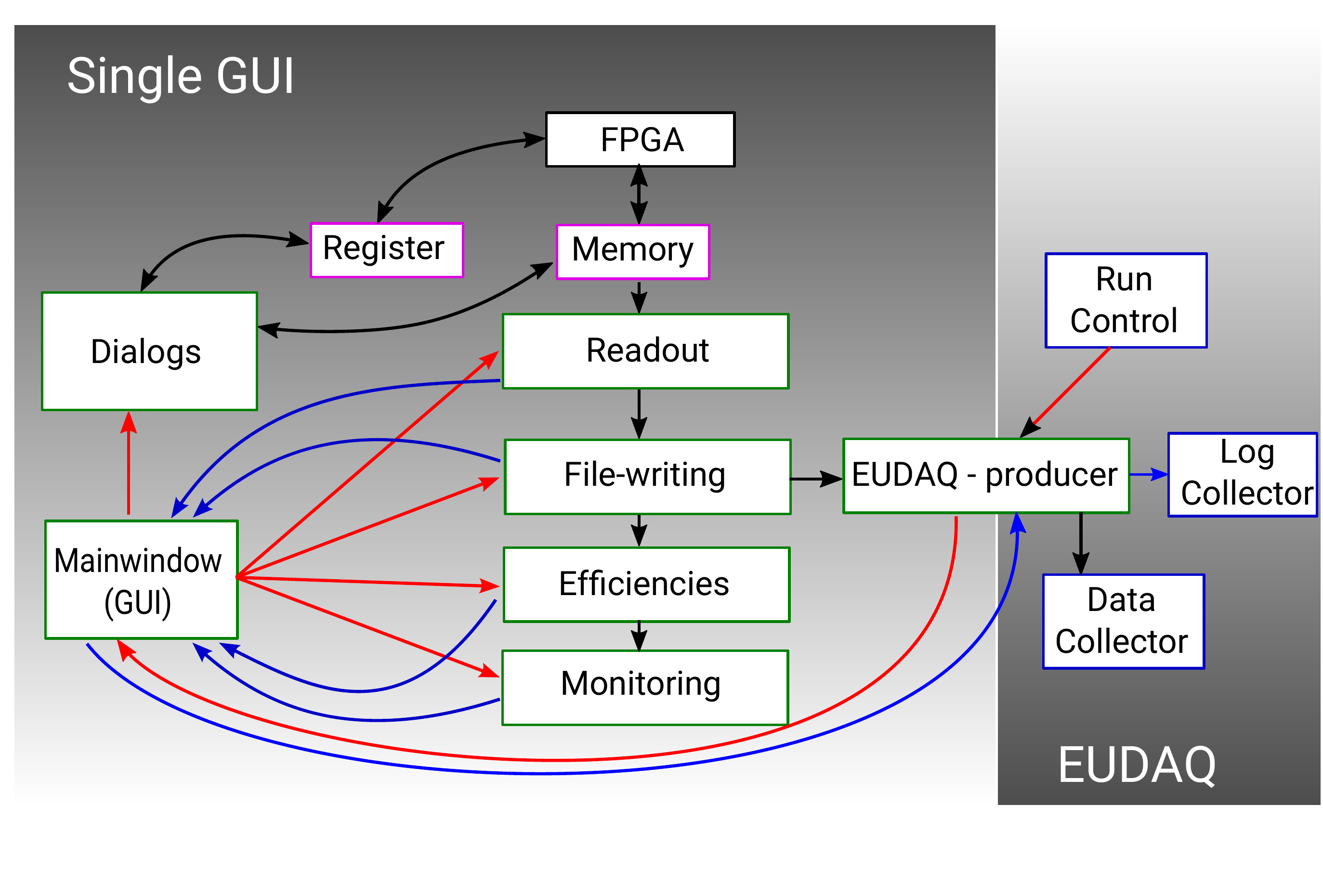}
	\caption{\label{mu3e:integration} Block diagram of the \EUDAQ integrated {{MuPix}} DAQ. The \EUDAQ \Producer serves as an
	interface between the EUDAQ and the {{MuPix}} DAQ. The blue rectangles are \EUDAQ components. The green ones are
	threads within the {{MuPix}} readout and the magenta ones are \emph{memories}.
	The black arrows indicate data flows, the red ones steering and the blue ones status information. Figure from \cite{mu3e-huth}.}
\end{figure}

A custom \Producer, serving as interface between the {{MuPix}} DAQ and the \EUDAQ \RunControl, has been developed.
The flow chart of the interaction between the two systems is depicted in Figure~\ref{mu3e:integration}.
The \EUDAQ \Producer steers the complete MuPix readout and configuration procedure.
Error messages, status information and the start/stop/configure signals are forwarded to the MuPix DAQ.
The \Producer buffers the entire data stream of the MuPix until one TLU trigger is registered, transforms the data
into an \EUDAQ interpretable format and sends it to the \DataCollector.

%% file: content/conclusion.tex
\section{Conclusion \& Outlook}
\label{sec:conclusion}

\EUDAQ has been developed as a common data acquisition software for many different use cases and fulfills several common requirements:
\begin{itemize}
    \item Flexible and light-weight: \EUDAQ is platform independent and can be operated on Linux, Windows or macOS, and comes with
    as few external dependencies as possible.
    \item Easy-to use: \EUDAQ incorporates well-defined interfaces which allow data sending and control of the user's
    hardware and conversion of the data for monitoring or offline analysis.
    \item Robust data taking: The data taking strategy is to store detector-native raw binary data without additional processing by the DAQ system. This allows to retain all information initially recorded and reduces the risk of data loss through errors in the processing code.
	    The data handling is performed by one central instance and data from different sub-detectors are
	    synchronised on an event-by-event level, ensured by the global busy-trigger communication provided by the \EUDET TLU.
\end{itemize}

Thanks to many user groups continuing to share their code integrating their own devices, \EUDAQ now ships with support for many different devices.
This both serves as a reference for new users and allows developers to ensure compatibility when modifying core parts of the framework.
Over more than a decade, \EUDAQ has thus been maintained for production and, at the same time, been constantly developed and improved.
The diversity of the contributing user groups is a testament to the collaborative spirit of the test beam community and the central role
that the \EUDET-type beam telescopes, and in turn \EUDAQ, plays.
As part of the \EUDET-type beam telescopes, \EUDAQ was assessed by the community at the ``Beam Telescope and Test Beam'' workshop in 2018~\cite{bttb6forum}.
From the discussion and survey among the participants, it was concluded that the \EUDAQ architecture and development road-map fulfills the test beam user's needs for the foreseeable future.

During the implementation phase of the \AIDA framework~\cite{aida}, \EUDAQ became a candidate for a common test beam DAQ, extending beyond its initial goal of serving as system for telescope-based test beams only.
In a comparison with other DAQ frameworks such as XDAQ~\cite{Brigljevic:2003kg}, \EUDAQ was judged as ``greatly simplified, and relies on as few external libraries as possible. \EUDAQ is designed to be portable.''
Furthermore, recommendation for improvements were collected from the telescope community.
These included suggestions such as parallel data streams and the possibility to acquire individual events for each particle or trigger \cite{behrens2015}.

These requirements led to the development of an improved version of \EUDAQ and the new version -- called \EUDAQII~-- was selected as the common software framework under the work package WP5 within the \AIDAII  framework \cite{aida2020}.
\EUDAQII breaks with several concepts such as the event-by-event synchronization of devices and is therefore not compatible with \Producers written for \EUDAQ.
It comes with the possibility to run multiple \DataCollector instances for decentralised data taking and is therefore more scalable than \EUDAQ.
Together with the new \AIDA TLU~\cite{aidatlu_paper} a higher trigger rate can be achieved and events can be built either online or offline by synchronizing the individual data streams using their trigger IDs or a common clock.
The framework was released in 2017~\cite{liu2019} and has already been applied by the ATLAS ITk Strips test beam group for the first time.
Thank to these enhancements \EUDAQ has become even more flexible and is ready for testing the next generation of detectors for future experiments.

%% file: acknowledgment.tex
\acknowledgments

The authors are greatly indebted to their colleagues from the DESY FH division over the last decade, who are essential to successfully maintain and extend this software framework.
This work was supported by the Commission of the European Communities
under the 6$^{\rm th}$ Framework Programme ``Structuring the European Research Area'', contract number RII3-026126.
The research leading to these results had received funding from the European Commission under the FP7 Research Infrastructures project AIDA, grant agreement no. 262025.
This project has received funding from the European Union's Horizon 2020 Research and Innovation programme under Grant Agreement no. 654168.
The support is gratefully acknowledged.
The information herein only reflects the views of its authors and not those of the European Commission and no warranty expressed or implied is made with regard to such information or its use.
